\documentclass{aa}  
\usepackage{graphicx}
\usepackage{xcolor}
\usepackage{txfonts}

\begin{document}

   \title{Recoiling Black Holes I. }\subtitle{Burst of Observables}

   \author{E. Hochart\inst{1} \and
          S. Portegies Zwart\inst{1}}
   \institute{
             Leiden Observatory, University of Leiden, 
             Niels Bohrweg 2, 2333 CA Leiden\\
             \email{hochart@mail.strw.leidenuniv.nl}
             }

   \date{Received 05/06/2025; accepted 29/07/2026}
 
  \abstract
   {}
   {To investigate the tidal disruption and gravitational wave events shortly after a massive black hole binary merges in the galactic centre and whose remnant black hole is ejected from the galaxy.}
   {Using computational methods, black holes of mass $M_{\bullet}=10^{5}$ M$_\odot$ and $4\times10^{5}$ M$_\odot$ embedded in a nuclear star cluster are kicked at velocities of $v_k=300$ and $600$ km s$^{-1}$. Systems are integrated for $0.1$ Myr using a $4$th-order Hermite scheme.}
   {The kick instantaneously repopulates the loss cone, producing a strong burst of tidal disruption events and gravitational wave mergers. The anisotropy in the apsidal orientation of bound stars prolongs this burst phase. Rates increase for lower $v_k$, larger $M_{\bullet}$ and for steeper nuclear star cluster density profiles at moment of merger.}
   {Assuming binary black holes scour a Bahcall-Wolf density profile during coalescence, ejected remnants with mass between $10^{5}\leq M_{\bullet}$ [M$_\odot] \leq 4\times10^{5}$ generate observable offset events at a forecasted rate of $\dot{N}\lesssim290$ yr$^{-1}$ up to redshift $z=3$. If, at the moment of merger, the milliparsec scales of the nuclear star cluster are described by a shallow density profile ($\gamma=1$), this decreases to $\dot{N}\lesssim30$ yr$^{-1}$. The strong dependence on the initial density profile and recoil kick makes observables powerful probes of the nuclear star cluster post-massive black hole binary coalescence, and provides test for numerical relativity.}


     \maketitle

\section{Introduction}
    The hierarchical growth of structure dictated by $\Lambda$CDM cosmology implies galaxy-galaxy mergers as common place events \citep{1978MNRAS.183..341W}. Observations and theory indicate that supermassive black holes (SMBH, with mass $M_{\bullet}\gtrsim 10^{6}$ M$_\odot$) often reside in galactic cores \citep{1995ARA&A..33..581K, 2005SSRv..116..523F}. Considering dynamical systems tend towards equipartition post galaxy-galaxy mergers SMBH binaries (SMBHB) readily form \citep{1980Natur.287..307B}. If they achieve small enough separations, gravitational wave (GW) emission instigates a merging event.

    Asymmetry in the binary's mass, spin amplitude and spin orientation leads to anisotropic emission of GWs. As a result, to conserve linear momentum, the newly formed SMBH will receive a kick \citep{1962PhRv..128.2471P, 1973ApJ...183..657B}. Numerical relativity show that these kicks can at times exceed velocity's of $v_{k}>10^{3}$ km s$^{-1}$ \citep[i.e][]{2007PhRvL..98w1102C}. Since SMBHB mergers occur in dense stellar environments, the recoiling remnant may carry a retinue of stars and compact objects forming a `hyper-compact stellar cluster' (HCSC). Initially, the HCSC extends a few milliparsecs in size \citep{2004ApJ...613L..37B, 2009ApJ...699.1690M} and roughly consists of particles (i.e stars, black holes, neutron stars...) whose orbital velocity at moment of merger exceeds the recoil kick. Such systems may also emerge from triple interactions \citep{2021PhRvD.104h3020B}. Although there have been tantalising signs of their detection, none are confirmed \citep[i.e,][]{2005Natur.437..381M, 2010MNRAS.407..645J, 2023ApJ...946L..50V, 2512.00174B}.
    
   Using numerical simulations, \citet[][hereafter KM08]{2008ApJ...689L..89K} predicted the rate of tidal disruption events (TDEs) from recoiling SMBHs kicked with large recoils. \citet{2012MNRAS.422.1933S} extended on this work and generalised their results by exploring a wider range in parameters. Here, we revisit the problem and focus on the phase shortly after recoil. Here, we include a stellar mass distribution, allowing analysis of potential extreme mass ratio inspiral (EMRIs) events. 
   
   With these systems theoretically abundant and the James Webb Space Telescope, LISA and the Vera C. Rubin Observatory capable of observing such systems \citep{2017arXiv170200786A, 2019ApJ...873..111I}, the investigation here employs direct $N$-body simulations and extends on previous research. Simulations are restricted to intermediate-mass black holes (IMBHs). These black holes (BHs) may be hosted in dwarf galaxies and/or represent the seeds of SMBHs \citep{2001ApJ...562L..19E, 2002ApJ...576..899P, 2006MNRAS.370..289B, 2013ApJ...775..116R, 10.1088/2041-8205/809/1/l14, doi:10.1142/S021827181730021X}. Results are also extrapolated to larger BH masses, although future work should be conducted to see if this is reasonable.
    
\section{Initial conditions}
    \subsection{Theoretical considerations}\label{Sec:Theory}
        Dynamical friction causes the most massive particles to settle into the centre of astronomical environments. In the case of galaxy-galaxy mergers, if both galaxies host an SMBH, dynamical friction will eventually form an SMBHB.  The process is efficient down to separations of about a parsec. Beyond this point the binary's evolution stalls because the nearby stellar population capable of exchanging energy with either constituents has been depleted. This is traditionally known as the final parsec problem. 
        
        In the last decade, research has found that the departure from spherical potentials allows for centrophilic orbits, resulting in an efficient repopulation of the loss cone and alleviating the final parsec problem \citep[i.e,][]{2006ApJ...642L..21B, 2013ApJ...773..100K, 2015ApJ...810...49V}. High-resolution numerical simulations find that this also holds for IMBHB embedded in replica dwarf galaxies \citep{2021MNRAS.508.1174K}.
        
        Upon further binary evolution, if the SMBHB semi-major axis, $a$, falls below some critical value estimated as, 
        \begin{equation}
            a\lesssim a_{\rm GW}\approx2\times10^{-4}r_{\bullet}, \label{Eqn:aGW}
        \end{equation}
        then GW emission induces the SMBHB to inspiral \citep{2007ApJ...671...53M}. Here $r_{\bullet}$ is the BH sphere of influence, 
        \begin{equation}
            r_{\bullet}\approx \frac{GM_{\bullet}}{\sigma^2}, \label{Eqn:rinfl}
        \end{equation} 
        with $G$ being the gravitational constant and $\sigma$ the bulge velocity dispersion. Observations indicate the latter has a tight correlation with the SMBH's mass, $M_{\bullet}$, scaling as \citep{2005SSRv..116..523F},
        \begin{equation}             
            \frac{M_{\bullet}}{10^8{\rm M} _\odot} = 1.66\left(\frac{\sigma}{200{\rm km s}^{-1}}\right)^{4.86}. \label{Eqn:FerrareseFord}
        \end{equation}
        This correlation appears to hold at $M_{\bullet}\sim10^{5}$ M$_\odot$  \citep{2006ApJ...641L..21G, Schutte_2019, 2020ApJ...898L...3B}.
        
        If mass or spin asymmetry exists within the SMBHB, the merger remnant receives a kick whose velocity can range from $10$ to $10^{3}$ km s$^{-1}$ \citep[i.e,][]{2008ApJ...682L..29B, 2015PhRvD..92b4022Z}. This paper fixes recoil kicks to $v_{k} = 300$ km s$^{-1}$ and $600$ km s$^{-1}$. Lower kicks were not considered since the paper focuses on systems offset from the galactic nuclei which they originate from. The radius of the recoiling system is \citep{2007ApJ...671...53M},
        \begin{equation}
            r_{k} \lesssim \frac{8GM_{\bullet}}{v_{k}^2}. \label{Eqn:Rkick}
        \end{equation} 
        Given the high densities in nuclear star clusters (NSC), the recoiling SMBH can carry a retinue of particles as it gets ejected from the nuclei and, while scattering events between the SMBHB and field stars depletes the surrounding area, two-body relaxation acts to repopulate it by supplying particles whose periastron exceeds the semimajor axis of the SMBHB. 
        
        Once the binary's semi-major axis falls below $a_{\rm GW}$, it decouples from the environment, and two-body relaxation can no longer efficiently replenish the population, resulting in a cavity at $r\lesssim a_{\rm GW}$ \citep{2007ApJ...671...53M}. Assuming a power-law density distribution, at moment of merger, the environment has \citep{2009ApJ...699.1690M},
        \begin{equation}
            \rho(r) =  \left\{\begin{array}{lr}
            0 & \text{for } r\lesssim a_{\rm GW},\\
            \rho_0\left(\frac{r}{r_0}\right)^{-\gamma} & \text{for } a_{\rm GW}\lesssim r \lesssim r_{\bullet}, \\
            \rho_0 & \text{for } r_{\bullet} \lesssim r.
            \end{array}  \right.\label{Eqn:Density_Distr}
        \end{equation}
        The profile depends on whether the environment is gas-poor (`dry') or gas-rich (`wet'), and if the NSC is in the collisional ($M_{\bullet}\lesssim 10^{8}$ M$_\odot$) or collisionless ($M_{\bullet}\gtrsim 10^{8}$ M$_\odot$) regime. 
        
        Massive, evolved, galaxies typically exhibit shallow central density profiles with $0.5\lesssim\gamma\lesssim1.5$ \citep{2001ApJ...563...34M, 2006ApJ...648..976M, 2009ApJ...699.1690M}. Contrastingly, lower-mass galaxies are collisional and thus more efficiently repopulate the central regions. As a result, they maintain a steeper profile, with $\gamma = 1.75$ \citep{2001ApJ...563...34M, 2006ApJ...648..976M, 2007ApJ...671...53M, 2009ApJ...699.1690M}. 
        
        In gas-rich mergers, the presence of gas shortens the SMBHB merger timescale, mitigating the impact of core scouring such that $\gamma$ remains large, however the enhanced merger timescale also allows less time for relaxation mechanisms to refill the depleted region.  In addition, star formation can occur within the gas-rich environment, further enhancing $\gamma$ \citep{2009ApJS..182..216K}. Analysis of these competing effects remain outside the scope of the paper. Numerical simulations conducted within this investigation consider the collisional, dry regime, assuming recoiling BHs with mass $M_{\bullet}=10^{5}$ M$_\odot$ and $4\times10^{5}$ M$_\odot$. This choice also results in us using $\gamma =1.75$ \citep[Bahcall-Wolf cusp,][]{1976ApJ...209..214B, 2007ApJ...671...53M}. 
        
        Equation~\ref{Eqn:Density_Distr} is a first-order approximation of the density profile around a remnant BH in the galactic nuclei. During inspiral, an SMBHB may influence the morphology of the environment in ways not considered here. Recent studies suggest that, on parsec scales, the NSC may be oblate at moment of merger due to the interactions between the SMBHB and background particles \citep{2023MNRAS.521.6089M, 2025A&A...693A..22M}. While such flattening could, in principle, extend down to the $\sim10$ mpc scales relevant to this work, a spherical distribution is adopted. This assumption is justified given the relatively coarse mass resolution ($M = 152.6,{\rm M_\odot}$) in those simulations, which make direct extrapolation to sub-parsec scales uncertain. It is also worth noting that their simulations end once the SMBHB is hard and do not resolve the final inspiral. Since the inspiral timescale at this semi-major axis still exceeds the local relaxation timescale, it hinders the accuracy of the NSC at mpc scales.
        
    \subsection{Nuclear star cluster initialisation}\label{Sec:NSC_IC}
        Stars are initialised with masses between $M_{*}\in[0.5, 100]$ M$_\odot$, sampled from a power-law distribution with $\alpha = -1.35$, flatter than a Salpeter $\alpha=-2.35$ \citep{1955ApJ...121..161S}. This initial mass function (IMF) follows from observations of the Milky Way galactic centre and numerical work on star formation in the inner parsec of NSCs \citep{2003ApJ...594..812G, 2012ApJ...749..168M}. To account for an older stellar population, particles are evolved to $100$ Myr using \texttt{SeBa} \citep{1996A&A...309..179P, 2001A&A...365..491N} before any integration is done.
        
        The choice of pre-evolving stars to $100$ Myr is motivated by observations and generalisability. At $100$ Myr, massive stars have long reached the final stages in their life cycle, therefore, there is little scatter in demographic between stellar populations beyond $t\gtrsim100$ Myr. The same cannot be said if one were to model younger stellar populations of $\sim1-10$ Myr. Additionally, observations of the Milky Way NSC indicate that $\sim85\%$ of stars have ages $\gtrsim10$ Gyr, $\sim15\%$ have ages $\lesssim3$ Gyr and only a few percent below $100$ Myr \citep{2020A&A...641A.102S}. Taking this proportion as a roughly universal trend, NSCs are dominated by old stars. Lastly, an older population better represents the dry merger scenario considered here, since the lack of gas suppresses star formation.
        
        The assumed IMF and initial stellar ages influence the results. For fixed NSC density, a bottom-heavy IMF contains more stars than a top-heavy IMF. As a result, the emerging HCSC will be more populated and exhibit higher TDE and GW rates (cf. Equation~\ref{Eqn:Gamma_Relation}). Due to stellar mass loss,  an older initial population for a fixed NSC density will also exhibit the same effect since the average stellar mass, $\langle m_*\rangle$, is reduced. 

        The NSC is initialised with \texttt{AGAMA} \citep{2019MNRAS.482.1525V}. \texttt{AGAMA} is a \texttt{C++} software library dedicated towards galaxy and stellar dynamics. Its action–angle formalism allows it to reconstruct stellar systems in equilibrium for a range of analytical potentials, making it an ideal tool to initialise stellar systems for $N$-body simulations. As mentioned in section \ref{Sec:Theory}, Bahcall-Wolf cusps are considered ($\gamma=1.75$, $f(E)\propto E^{1/4}$). In all cases, the stellar cusp is centred around the IMBH and following Equation~\ref{Eqn:Density_Distr}, particles with periastron $r_p<a_{\rm GW}\approx2\times10^{-4}r_{\bullet}$ relative to the central IMBH get removed. 
        
        This approach underestimates the region devoid of stars in reality, and thus overestimates the HCSC compactness. In turn, this increases predicted TDE and GW rates. Indeed, simulations indicate that binary MBH eccentricities are strongly stochastic \citep{10.1093/mnras/stac241, Rantala_2024}. Given the steep $t_{\rm GW}\propto(1-e^2)^{7/2}$ dependence of GW inspiral \citep{1963PhRv..131..435P}, higher eccentricities shorten the inspiral, reducing the time available for two-body relaxation to repopulate the depleted region.
        
        Once setup, the central IMBH receives a kick. The orbital eccentricity, $e$, for each star with respect to the IMBH is computed, keeping track of those with $e<1$ since these constitute the bound population which eventually form the HCSC. Typically, the HCSC mass agrees with the predictions of \citet{2009ApJ...699.1690M} (Equation~\ref{Eqn:MHCSC}) to $\pm1\%$,
        \begin{equation}
            M_{\rm HCSC} = F_1(\gamma)M_{\bullet}\left(\frac{GM_{\bullet}}{v_{k}^2 r_{\bullet}}\right)^{3-\gamma} \label{Eqn:MHCSC},
        \end{equation}
        with $F_1(\gamma)$ being a numerical factor found as $F_1(\gamma)=11.6\gamma^{-1.75}$ \citep{2009ApJ...699.1690M}.

    \begin{table}
        \caption{Summary of simulated configurations.}
        \label{Tab:IC} 
        \centering 
        \begin{tabular}{c c c c}
            \hline\hline
                $M_{\mathrm{IMBH}}$ [M$_\odot$] & $N_{\rm Run}$ & $\gamma$ & $v_{\rm k}$ [km s$^{-1}$] \\ \hline 
                $10^{5}$ & $5$ & $1.75$ & $300$ \\
                $10^{5}$ & $10$ & $1.75$ & $600$ \\
                $4\times10^{5}$ & $5$ & $1.75$ & $300$ \\
                $4\times10^{5}$ & $5$ & $1.75$ & $600$ \\
            \hline          
        \end{tabular}
        \tablefoot{Col. 1: IMBH mass. Col. 2: Number of realisations simulated. Col. 3: Initial cluster density profile power-law. Col. 4: IMBH kick velocity.}
    \end{table}
        
    \subsection{Numerical set-up}\label{Sec:Methods}
        When $v_{k}=600$ km s$^{-1}$, $M_{\bullet}=10^{5}$ M$_\odot$, ten direct $N$-body simulations are conducted until they reach $0.1$ Myr. For the other combination of parameters five runs are done to the same end point\footnote{For a movie, see: https://www.youtube.com/shorts/5vdK4-7W-H4}. Each realisation has a unique set of initial conditions since the random seed ensures that different masses and phase-space coordinates are sampled from the underlying IMF and stellar energy distribution (see section \ref{Sec:NSC_IC}). Assuming stars orbit with semi-major axis $a\approx r_{k}$, $0.1$ Myr covers between $5\sim180$ orbits.
        
        The code uses the Astronomical Multipurpose Software Environment (\texttt{AMUSE}) \citep{2009NewA...14..369P, 2013A&A...557A..84P, 2018araa.book.....P}, which consists of numerous codes spanning different physical regimes such as hydrodynamics, stellar evolution or gravitational integration. Here, the 4th-order predictor-corrector Hermite scheme \texttt{Ph4} \citep{2012ASPC..453..129M} with internal time-step parameter $\eta = 0.1$ is used to evolve the system. No softening is applied. 
        
        \texttt{Ph4} solves the Newtonian equations of gravity, omitting relativistic corrections. While this modifies the collisional history of our system, likely suppressing the final rate \citep{10.1093/mnras/sty197, 2021ApJ...907L..20T}, we adopt this approximation for computational efficiency. Indeed, including post-Newtonian (PN) terms up to $2.5$PN order, which account for GW emission, would increase the scaling of computational operations from $\mathcal{O}(N^2)$ to $\mathcal{O}(N^3)$.

        An additional consequence of us evolving the system under classical gravitational laws is that we do not explicitly model GW-driven inspirals. To still allow for their analysis, we identify collisions of compact objects with the central MBH as proxies. Compact objects are identified automatically with \texttt{SeBa}. From this subset of collisional events we exclude direct plunges by considering only those events whose periastron distance with the MBH exceeds the MBH's Schwarzschild radius. The remaining events correspond to compact object whose trajectory enter the relativistic capture regime and have potential to develop into EMRIs if relativistic effects were included, and thus are classified as potential EMRI progenitors.

        This classification should not be confused with a self-consistently modelled EMRI. Genuine EMRI's consist of a compact object bound to a massive black hole that gradually loses orbital energy and angular momentum through GW emission, completing many orbits before its final plunge. However, since our simulations omit PN terms, we cannot determine whether identified progenitor's would indeed undergo an inspiral, experience a head-on plunge, or be perturbed out of the capture region. The events identified here therefore represent possible EMRI progenitors rather than confirmed EMRIs. Considering results from \citet{2024A&A...685A.123H}, our use of a classical versus a PN algorithm likely means we underestimate the number of EMRI events alongside their final strains.
        
        Stellar evolution is considered using \texttt{SeBa} \citep{1996A&A...309..179P, 2012ascl.soft01003P}, with data on the stellar radii and mass being transferred to the gravitational integrator every $100$ yr.

        \subsection{Collision detection}\label{Sec:Colls}
        Once two particles are separated by a distance less than their combined radii, they merge. Collisions are resolved using the `sticky sphere' approximation, such that the system's mass and momentum are conserved. Particle types are identified by \texttt{SeBa}, which influences their collisional radii.
        
        BHs have their collision radius equal their innermost stable circular orbit (ISCO), $R_{\rm ISCO} = 6GM/c^2$, with $c$ the speed of light. Neutron stars (NS) have their radius taken from \texttt{SeBa}. For GW events, the remnant receives a recoil kick using equation (3) of \citet{2012PhRvD..85h4015L} with an additional $(1+e)$ factor to account for any residual eccentricity \citep{2008ApJ...686..829H}. By assuming spinless particles, these kicks are underestimated. 

        Stars and white dwarfs have their radius equal their respective tidal radius relative to the central IMBH. That is \citep{1988Natur.333..523R},
        \begin{equation}
            R_{\rm tide} = R_{*}\left(\eta^2\frac{M_{\bullet}}{m_{*}}\right)^{1/3} \label{Eqn:Rtide},
        \end{equation}
        with $\eta=0.844$ for an $n=3$ polytrope \citep{1995MNRAS.275..498D}. If a star collides with a compact particle that is not the IMBH, the tidal radius is temporarily scaled by a factor $(M_{\rm c.o}/M_{\bullet})^{1/3}$ and checked whether the separation lies below this new tidal radius. In the case of a non-TDE event (star-star mergers or a WD merging with a non-IMBH compact object), the radius is temporarily changed to those provided by \texttt{SeBa} and the code checks whether their separation falls within their combined radius.

\section{Results}
        \subsection{Loss cone repopulation}
        \subsubsection{Event rate}
            \begin{figure*}
                \centering
                \includegraphics[width=.42\textwidth]{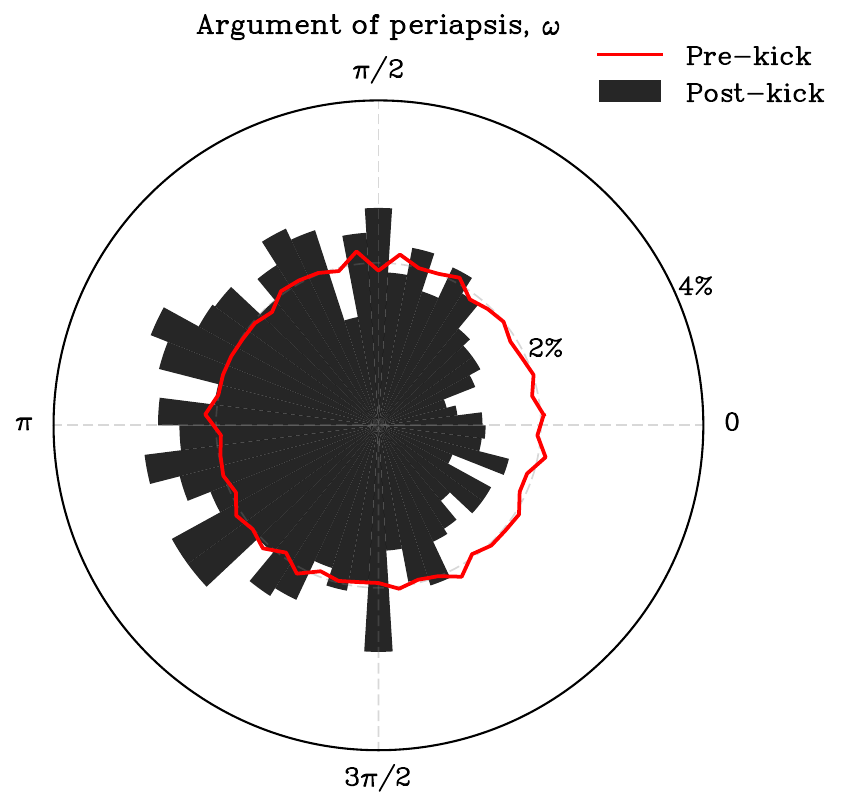}
                \includegraphics[width=.42\textwidth]{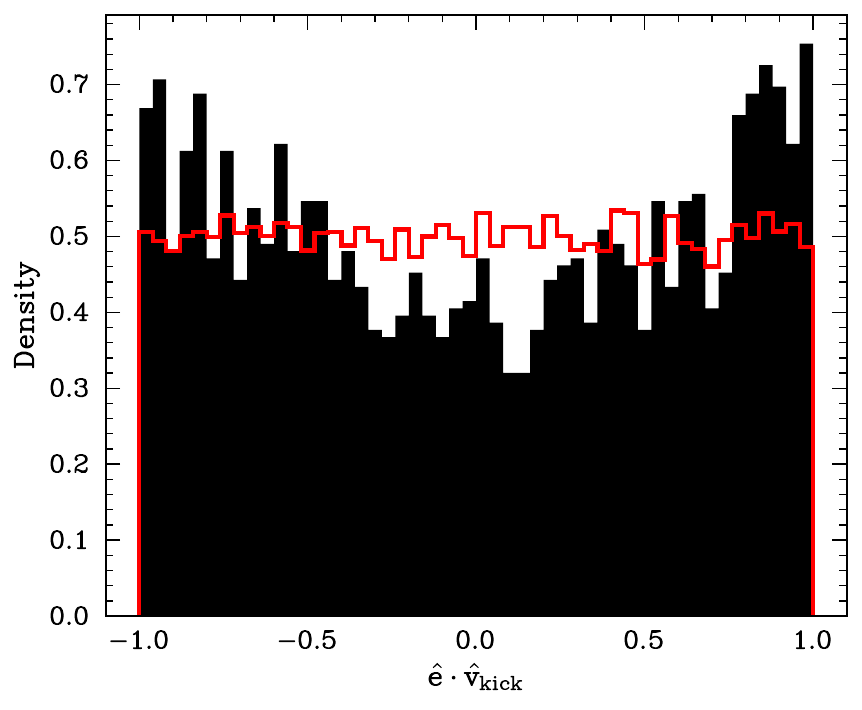}
                \caption{Left: Particle argument of periapsis, $\omega$, pre-kick (red contour) and post-kick (black bars). Right: Normalised distribution of the eccentricity-vector, $\hat{e}$, orientation relative to the kick direction, pre-kick (red) and post-kick (black). Data considers $M_{\bullet}=4\times10^{5}$ M$_\odot$ and $v_{k}=600$ km s$^{-1}$. }
                \label{Fig:ApsidalAlignment}
            \end{figure*}
            At the moment of a recoil kick, all particles experience a velocity translation, $-v_k$, relative to the recoiling BH. This impulsive change alters the orbital angular momenta of the particles with some now being within the recoiling BH's loss cone, $L_{\rm min}\approx\sqrt{2GM_{\bullet}R_{\rm tide}}$. In a triaxial system, as the case for a HCSC (see Appendix A), the fraction of particles of energy, $E$, that lies within the loss cone goes as \citep{2013CQGra..30x4005M},
            \begin{equation}
                P_{\rm LC}(a)\sim\sqrt{\frac{(1-q)}{2}}\frac{L_{\rm min}}{L(E)}, \label{Eqn:FracLoss}
            \end{equation}
            where $L(E)\approx \langle m_*\rangle v_k a$ and $q$ is the ratio between length of the short and long axis.
            
            In addition to this impulsive loss-cone filling, the kick generates an anisotropy within the recoiling system. The left-hand panel of Figure~\ref{Fig:ApsidalAlignment} shows the distribution of the particle argument of periapsis, $\omega$, before and after the kick. While the pre-kick distribution is approximately isotropic, the post-kick becomes polarised towards $\pi/2 \leq \omega \leq 3\pi/2$, indicating a preferred orbital geometry \citep[see also][]{2018ApJ...853..141M, 2021ApJ...921L..12A}. Meanwhile, the right-hand panel shows the distribution of $\mu_e = \hat{\mathbf{e}}\cdot\hat{\mathbf{v}}_{\rm k}$,
            where $\hat{\mathbf{e}}$ is the unit eccentricity vector and $\hat{\mathbf{v}}_{\rm k}$ is the unit kick direction. The presence of peaks at $\hat{\mathbf{e}}\pm1$ indicate that bound particles tend to lie on the plane parallel to the recoil kick, with the negative and positive sign being near symmetric and denoting alignment and anti-alignment with the kick direction. The plot shows once more the tendency for the bound system to be preferentially aligned.
            
            In apsidally aligned system, stars that drift ahead of the aligned group experiences a torque that reduces its angular momentum while keeping its orbital energy nearly constant. Conversely, stars lagging behind it will gain angular momentum. These alternating torques drive oscillations in eccentricity that repeatedly repopulate the loss cone and can prolong the initial burst of tidal disruptions and merging events for $\gtrsim10^{4}$ orbital periods. 
            
\begin{figure*}
                \centering
                \includegraphics[width=.98\textwidth]{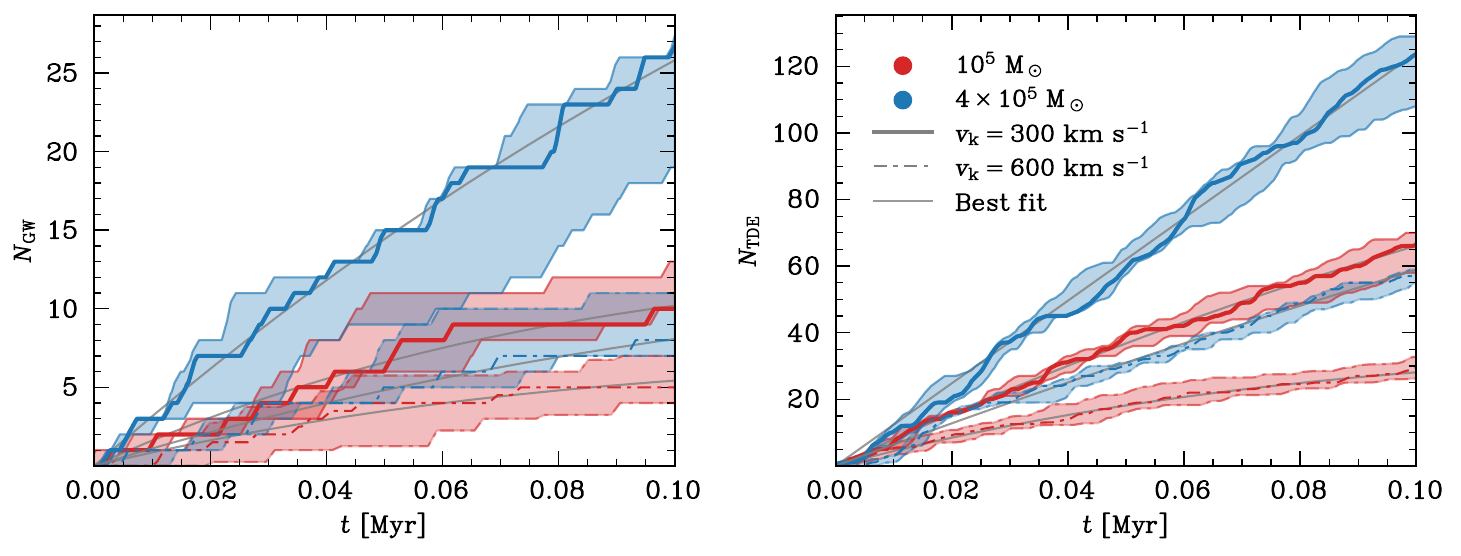}
                \caption{Collision events with the central IMBH in time. Left corresponds to GW events while right, TDEs. Solid coloured lines represent the median events when $v_{k} = 300$ km s$^{-1}$, while dotted lines $v_{k}=600$ km s$^{-1}$. Shaded regions in $75\%$ and $25\%$ interquartile range. Gray lines represent the best-fit (Equation~\ref{Eqn:Gamma_Fit}). }
                \label{Fig:ALL_Events}
            \end{figure*}

            This resonant torque acts on precession timescales \citep{2018ApJ...853..141M, 2024ApJ...966L...4A},
            \begin{equation}
                t_{\omega} \approx \frac{M_{\bullet}}{M_{\rm HCSC}}P_{\rm orb}(a). \label{Eqn:t_omega_body}
            \end{equation}
            For the configurations considered here, $t_{\omega}$ ranges between $0.045 - 0.125$ Myr, with the shortest time scale corresponding to $M_{\bullet}=10^{5}$ M$_\odot$, $v_k=600$ km s$^{-1}$. This behaviour is reflected in Figure~\ref{Fig:ALL_Events} which shows the median over runs in the number of GWs (left) and TDEs (right) events involving the central IMBH in time. 
            
            The initial burst phase is observed to last for several orbital periods, and a damping in the best fit line appears by $t_{\rm sim}\approx50$ kyr in the configuration with the smallest $t_{\omega}$. Additionally, the oscillatory nature of the mechanism is hinted at within the plot, although the prominence of which is diluted due to ensemble averaging.
    
            The analytical event rate for $\gamma=1.75$ follows (Appendix A for derivation),
            \begin{equation}
                \Gamma(t)\propto A(\gamma) \frac{M_{\bullet}^{0.365}}{v_k\langle m_*\rangle} e^{-t/t_{\omega}}. \label{Eqn:Gamma_Relation}
            \end{equation}
            The coefficient $A(\gamma)$ is also derived in the appendix, but listed below for convenience,
            \begin{equation}
                A \equiv \frac{2\left(\left(2\gamma+3\right)\left(\left(2\gamma+1\right)-4\gamma+2\right)+4\gamma^2-1\right)}{\left(2\gamma-1\right)\left(2\gamma+1\right)\left(2\gamma+3\right)}.
            \end{equation}
            
            Removing all numerical factors within the extended expression (Equation~\ref{Eqn:FinalEqn}) and replacing them with coefficients $\alpha$, $\beta$ and $\eta$, the grey curves show solutions to the best fit trends shown in Figure~\ref{Fig:ALL_Events}, and follow the form,
            \begin{equation}
                \Gamma = \eta\cdot A\cdot B\cdot C\cdot D\cdot E^{-1} \cdot e^{-Dt}, \label{Eqn:Gamma_Fit}
            \end{equation}
            where,
            \begin{align}
                B &\equiv \sqrt{\frac{(3-\gamma)^2 (1-q)}{2}}G M_{\bullet}^2 \sqrt{R_{\rm tide}}, \\
                C &\equiv \frac{(8r_{\bullet})^{\gamma-3}}{v_k}a_{\rm GW}^{-(\gamma-1/2)}, \\
                D &\equiv \frac{\alpha\sqrt{GM_{\bullet}}}{\beta^{\gamma-3}r_{\bullet}^{3/2}}, \\
                E &\equiv \frac{M_{\bullet}^{1/3}\sqrt{GM_{\bullet}}}{M_{\rm HCSC}^{1/3}a_{\rm GW}^{3/2}}.
            \end{align}
            A non-spherical potential is essential since it provides the $1/v_k$ dependency via Equation~\ref{Eqn:FracLoss}, a trend observed in Figure~\ref{Fig:ALL_Events}. Instead, an isotropic system would have $P_{\rm LC}\sim(L_{\rm min}/L(E))^2\propto v_k^{-2}$. The corresponding best-fit parameters are $\eta\approx8.7\times10^{4}$, $\alpha\approx6.4\times10^{-4}$ and $\beta\approx9.55$. Scatter is present, particularly for $M_{\bullet}=4\times10^{5}$ M$_\odot$, partly due to simulations lasting $t_{\rm sim}\lesssim t_{\omega}$. 
            
            The coefficients encapsulate several physically motivated but not explicitly resolved dependencies. These include the total mass of the apsidally aligned population, the relation between $r_{\bullet}$ and the characteristic semi-major axis which sources the TDE and GW events, the ratio between that same characteristic semi-major axis with that of the clustered population, a factor defining the effective extent of the Hill radii and the mean eccentricity and angular dispersion of the eccentricity vectors within the aligned population, $e\phi_{\rm disk}$.
            
            One should caution that a limited range of $M_{\bullet}$ and $v_k$ were tested. We remind the reader that we only consider the case where $v_k>v_{\rm esc}$ to reduce computational time. Indeed, focusing on this regime allows us to neglect the effect of the galactic environment on the HCSC \citep[see i.e][]{2008ApJ...678..780G, 2012ApJ...748...65L}. With a similar motivation, only IMBH clusters are considered (the HCSC population going as $N_{\rm HCSC}\propto M_{\rm HCSC}\propto M_{\bullet}^{4-\gamma}$). As such, it isn't clear how tendencies will differ if we go beyond the tested regime; namely lower recoil kicks and larger BH masses. This is left for future work.

        \subsubsection{The dependence of $\gamma$}
           Results indicate a burst of EMRI candidates and TDEs occur shortly after a MBH merger event occurs. This burst phase can have merger rates exceed that in a static NSC given the right parameters. Indeed, when considering $v_k\approx v_{\rm esc}$, $M_{\bullet}=10^{5}$ M$_{\bullet}$ and $\gamma=2.0$, merging events are between $\sim30-100$ times greater than a case where the same BH mass is static in the galactic center (Figure~\ref{Fig:Event_vs_gamma}). For $\gamma=1.75$, this reduces to $7-20$. Extrapolating to larger BH masses ($M_{\bullet}=10^{6}$ M$_\odot$), then for $\gamma=2$, the increase is $\sim10^{2}-10^{3}$. 
           
           Figure~\ref{Fig:Event_vs_gamma} shows the strong dependence between the event rate and the initial cluster density profile, $\gamma$. If at the moment of merger, the surrounding region is characterised with density slope $\gamma\lesssim1.2-1.4$, then recoiling IMBH will appear more quiescent than if it were static. For $M_{\bullet}=10^{6}$ M$_\odot$, the corresponding thresholds are $\gamma\lesssim0.7-1.1$.
           
           \citet{2026A&A...706A.354K} find that IMBHB inspirals naturally develop compact-object cusps with $\gamma\simeq1.75$ within local relaxation times within the mpc regime. Similarly, \citet{2010ApJ...708L..42P} find through numerical simulations of two mass component systems around a MBH that cusps around $M_{\bullet}\lesssim5\times10^6$ M$_{\odot}$ BHs replenish within $\sim10-25\%$ of the local relaxation time, with the profile of massive components settling into $\gamma=2.1$, while the lighter components $\gamma=1.5$. Considering this, if such systems (and kicks) were common, they should be observed. However, \citet{2005LRR.....8....8M} show that major galaxy mergers have cusps settle with a density profile $\gamma\approx 1$. If this profile extends to the mpc region, then TDE and GW rates of any emerging HCSC will be lower than that of a static nuclei. Overall, future observations (or lack thereof) can reveal information of the density distribution of particles within mpc scales post-SMBHB mergers.
            
            While massive galaxies are predicted to have $\gamma\approx0.5$ \citep{2006MNRAS.367.1746M}, the analytical form of $A$ diverges at $\gamma=0.5$, indicating that the theory breaks down somewhere at lower $\gamma$. \citet{2009ApJ...699.1690M} note that no isotropic density distribution exists when $\gamma=0.5$. As a result, orbits will be mostly circular and the HCSC mass will start to increase. This would explain the upturn in rates at $\gamma\lesssim0.6$, but further analysis is required to confirm whether the trend truly holds at lower $\gamma$.

        \subsubsection{Resonant relaxation}
            Since particles within an HCSC are deeply embedded within the BH's gravitational potential, they follow near-Keplerian orbits. Such orbits maintain their toplogy over many orbital periods, and thus, constituents of the HCSC experience highly correlated encounters with one another. These highly correlated encounters result in the exertion of mutual torques in a process called resonant relaxation (RR), driving angular momentum diffusion on timescales \citep{1996NewA....1..149R, 1998MNRAS.299.1231R, 2006ApJ...645.1152H},
            \begin{equation}
                t_{\rm RR}\approx\frac{M_{\bullet}}{\langle m_*\rangle}P_{\rm orb}(a).
            \end{equation}
            This differs from the apsidal precession timescale, $t_{\omega}$ (Equation~\ref{Eqn:t_omega_body}), by a factor $M_{\rm HCSC}/\langle m_*\rangle$. Since $M_{\rm HCSC}\gg \langle m_*\rangle$, RR operates much slower, resulting in a reduced efficiency when repopulating the loss cone. For reference, the configurations tested here have $t_{\rm RR}\approx 0.027-3.5$ Gyr, with the timescale increasing for larger $M_{\bullet}$ and lower $v_k$. Since $t_{\rm RR}\gg t_{\rm sim}$, this mechanism doesn't imprint itself on runs here.
    
            This introduces an important, albeit subtle, effect on the results presented in KM08. In their study, the authors performed numerical simulations with integration times $t_{\rm sim} \gg t_{\omega}$, ensuring that RR effects come into play. This approach allowed them to extract the RR diffusion coefficient, $C_{\rm RR}$, a term which directly influences the rates of TDEs and GWs. The numerically determined $C_{\rm RR}$ was then inserted into their purely-RR driven analytical  expression to predict the event rate of recoiling clusters. While this method is well motivated, it implicitly assumes that RR drives TDEs and GWs even at the earliest stages, before the precession effects even come into play, and neglects the initial burst phase considered here. As a result, predicted rates are underestimated.
            
            \begin{figure}
                \centering
                \includegraphics[width=\columnwidth]{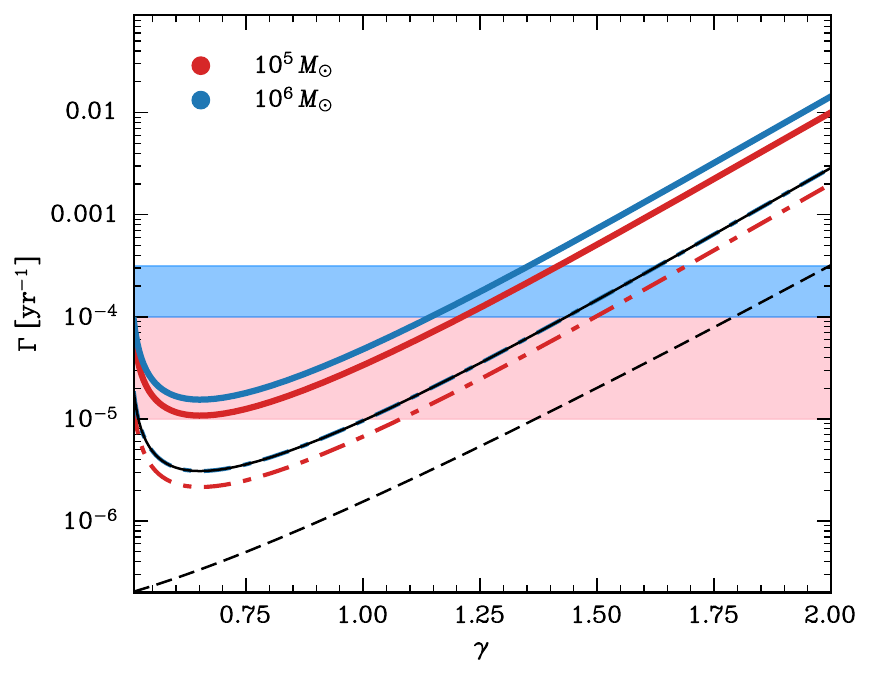}
                \caption{Thick lines use Equation~\ref{Eqn:Gamma_Fit} assuming $v_{\rm kick}=v_{\rm esc}$. Dashed lines assume $v_{\rm kick}=5 v_{\rm esc}$. In both cases we assume $v_{\rm esc}\approx2\sigma$. Note that this means $v_{\rm kick}$ will change depending on the BH mass. Black solid and dashed line fix $M_{\bullet}=10^{6}$ M$_\odot$ and $v_{k}=10^{3}$ km s$^{-1}$. This is to mimic a model of KM08. The solid line uses Equation~\ref{Eqn:Gamma_Fit}, while the dashed one equation 4 of KM08 where $r_{\bullet}=$ 3pc. The blue shaded region represents rates in galaxies hosting BHs of mass $M_{\bullet}=10^{5}\sim10^{6}$ M$_\odot$ \citep{ 2014ApJ...792...53V}. The red shaded region, those hosting BHS of mass $10^{7}\sim10^{8}$ M$_\odot$ \citep{2004ApJ...600..149W}.}
                \label{Fig:Event_vs_gamma}
            \end{figure}
            The underestimation is illustrated with the thin black solid and dashed curves of Figure~\ref{Fig:Event_vs_gamma}. Rates using Equation~\ref{Eqn:Gamma_Fit} are systematically larger by a factor $\sim5$ across $\gamma \gtrsim 0.7$. This particular example was chosen to match one of the best-fit lines in figure 2 of KM08. Decreasing $M_{\bullet}$ and $v_k$ will increase perceived differences between theories. 
            
            Analytically, a key difference emerges when realising that while Equation~\ref{Eqn:Gamma_Fit} scales as $v_k^{-1}$, RR who predicts rates to occur at a rate \citep[KM08]{2006ApJ...645.1152H}, 
            \begin{equation}
                \Gamma_{\rm RR}\approx C_{\rm RR}(\gamma) \frac{\ln\Lambda}{\ln R} \left(\frac{v_{k}}{r_{k}}\right) F_1(\gamma)\left( \frac{GM_{\bullet}}{r_{\bullet}v_{k}^2}\right)^{3-\gamma}, \label{Eqn:RR_Rate_Body}
            \end{equation}
            predicts $\Gamma \propto \sqrt{v_k}$ for $\gamma=1.75$, meaning that while greater kicks enhance rates in the RR regime, the opposite occurs when considering the burst phase. In the expression above, $\ln \Lambda$ is the Coulomb logarithm and $\ln R\approx\ln(r_k/R_{\rm tide})$. 
            
            To emphasise the difference, if one were to consider lower kicks, a $M_{\bullet}=10^{6}$ M$_\odot$ recoiling at $v_k=100$ km s$^{-1}$ yields $\Gamma_{\rm RR}\approx2.5\times10^{-5}$ yr$^{-1}$ when $\gamma=1.75$. Equation~\ref{Eqn:Gamma_Fit} would find $4.5\times10^{-3}$ yr$^{-1}$ instead. A difference of $180$. This underestimation of RR in resupplying the loss cone has strong implications observationally (see section \ref{Sec:Forecast}).

        \subsection{Observables}
            \subsubsection{Collisional events}
            Table \ref{Tab:EventQuant} summarises the types of events occuring for all four configurations. Generally, collisions with the IMBH form roughly $f\sim0.85$ of all events and a $1:6$ ratio between GW events and TDEs exist ($1:10$ when considering only events involving the IMBH). The $1:6$ GW-to-TDE ratio corresponds roughly with the $1:5$ ratio between compact and stellar objects in the simulations after initialisation. The ratio is smaller when considering only events containing the IMBH due to the dependence of tidal radius with mass. In a run which considered the NSC (but whose results are not considered here), no events occur with field (unbound) particles, agreeing with KM08 and \citet{2017MNRAS.467.4180S}. 
            
            Although the high densities in HCSCs likely enhance the production of blue stragglers, the rates are minimal. There are signs that the number of blue stragglers scales with $M_{\bullet}$ and inversely with $v_k$. This likely originates from either factor enhancing the HCSC population. A population of blue stragglers would rejuvenate the cluster, giving off an excess blue which could be used to identify distant unresolved HCSC. However,  \citet{2020MNRAS.495.1771L} find that caveats with use of a colour-colour diagrams exists unless the system exceptionally old and metal-rich. For instance, as blue stragglers age, they become bright giants which ages rather than rejuvenates the cluster. Indeed, when unresolved young ($t\approx1$ Gyr), sub-solar metallicity HCSC will appear as G- and F-stars in colour-colour diagrams, while older metal-rich systems ($t\approx10$ Gyr) occupy similar regions in colour-colour diagrams as K- and M-type stars. Stemming from galactic centers, the latter case is expected to happen the most. 
            
            In summary, the use of colour-colour diagrams is not the most reliable way in delineating HCSC and the current best known method in identifying such systems beyond the enhanced TDE and GW rates outlined in the previous section is their large mass-to-light ratios and the velocity dispersion (see also \citet{2009MNRAS.395.2127O, 2025ApJ...991...83R}).
            
            \subsubsection{Gravitational wave events}
            \begin{table}
                \caption{Rates of different collision events. The values represent the median, with $\pm$ values corresponding to the $25$th and $75$th percentile.}
                \label{Tab:EventQuant} 
                \centering 
                \begin{tabular}{c c c c c}
                    \hline\hline
                       & Config. 1 & Config. 2 & Config. 3 & Config. 4  \\\hline 
                        $N_{\rm GW}$ & $10^{+3}_{-0}$ & $5^{+2}_{-1}$ & $27^{+0}_{-8}$ & $8^{+3}_{-1}$ \vspace{0.15em} \\
                        $f_{\rm GW, \bullet}$ & $0.50^{+0.00}_{-0.00}$ & $0.50^{+0.25}_{-0.10}$ & $0.37^{+0.00}_{-0.00}$ & $0.50^{+0.25}_{-0.25}$ \vspace{0.15em} \\
                        $N_{\rm TDE}$ & $66^{+4}_{-8}$ & $29^{+4}_{-3}$ & $124^{+5}_{-16}$ & $57^{+2}_{-2}$ \vspace{0.15em} \\
                        $f_{\rm TDE, \bullet}$ & $0.97^{+0.00}_{-0.13}$ & $0.86^{+0.13}_{-0.08}$ & $0.82^{+0.09}_{-0.05}$ & $0.88^{+0.01}_{-0.07}$ \vspace{0.15em} \\
                        $N_{**}$ & $0^{+1}_{-0}$  & 
                        $0.5^{+0.5}_{-0.5}$ & $7^{+0}_{-3}$ & $3^{+1}_{-1}$ \\\hline  
                \end{tabular}
                \tablefoot{Col. 1: $M_{\bullet}=10^{5}$ M$_\odot$, $v_{k}=300$ km s$^{-1}$. Col. 2: $M_{\bullet}=10^{5}$ M$_\odot$, $v_{k}=600$ km s$^{-1}$. Col. 3: $M_{\bullet}=4\times10^{5}$ M$_\odot$, $v_{k}=300$ km s$^{-1}$. Col. 4: $M_{\bullet}=4\times10^{5}$ M$_\odot$, $v_{k}=600$ km s$^{-1}$. Row 1: The number of GW events occuring. Row 2: The fraction of GW events containing the central IMBH. Row 3: The number of TDEs. Row 4: The fraction of TDEs containing the central IMBH. Row 5: The number of mergers containing only stars and/or white dwarfs.}
            \end{table}
    
            \begin{figure}
                \centering
                \includegraphics[width=\columnwidth]{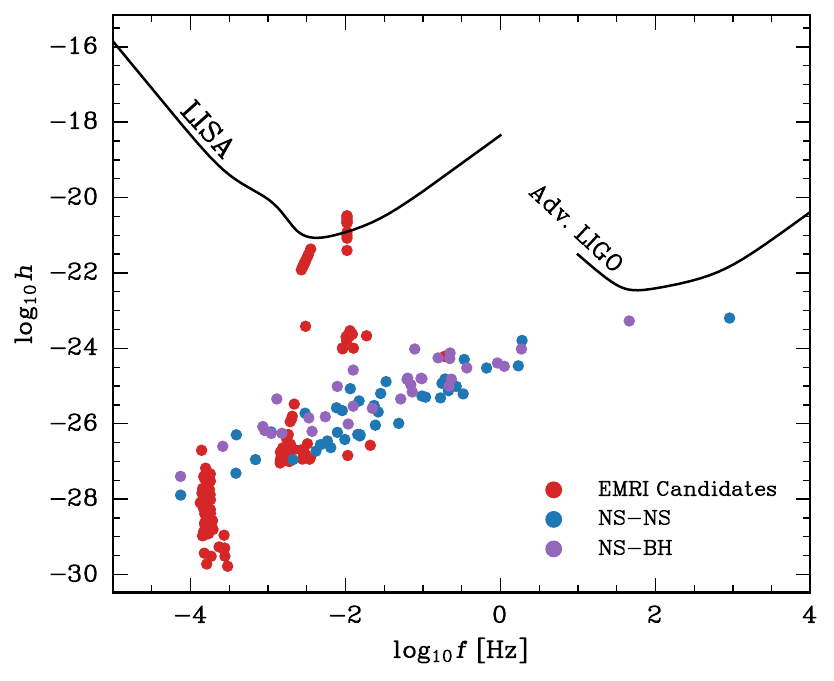}
                \caption{Frequency vs. strain for GW events occuring in all runs with $M_{\bullet}=4\times10^{5}$ M$_\odot$. Events are assumed to be sourced at redshift $z=1$ and adopt the cosmological parameters observed by \citet{2020A&A...641A...6P}. Orbital parameters are the osculating elements calculated at the last internal convergent time-step of the direct $N$-body code. Due to the lack of relativistic effects in this simulation, EMRI candidates are only representative (Section~\ref{Sec:Methods}). The strain is likely underestimated and computed from their osculating orbital elements at the final converged internal integrator time-step. LISA sensitivity curves are taken from \citet{2019CQGra..36j5011R} and Advanced LIGO from \citet{alex_nitz_2024_10473621}.}
                \label{Fig:GW_FreqStrain}
            \end{figure}
            
            Figure~\ref{Fig:GW_FreqStrain} shows the frequency-strain diagram of GW events occuring in all runs considering $M_{\bullet}=4\times10^{5}$ M$_\odot$. Calculations assume a redshift distance of $z=1$. For the EMRI candidates, the frequency was computed using equation~37 of \citet{2003ApJ...598..419W}, while the strain, equation~7 of \citet{2019PhRvD..99f3003K} (see also \citet{1963PhRv..131..435P, 2007ApJ...665L..59W, 2018MNRAS.481.4775D}). All other mergers were hyperbolic, producing gravitational bremsstrahlung \citep{1970PhRvD...1.1559P, 1977ApJ...216..610T, 1978ApJ...224...62K}. Gravitational bremsstrahlung is characterised by a transient burst of radiation, whose emission peaks at frequencies equal the reciprocal of their interaction time \citep{2006ApJ...648..411K}. Their strains are calculated following \citet{2018PDU....21...61G}. 
            
            Galactic nuclei are expected to be rich in gravitational bremsstrahlung sources, as fly-by interactions between compact objects naturally generate such signals \citep[i.e,][]{2008MPLA...23...99C, 2021MNRAS.506.1665G}. The plot only considers events ending in head-on collisions since these yield the most pronounced signals and suffice for analysis. In most cases even such violent events will remain undetectable for next-generation GW interferometers unless they occur in the local Universe (luminosity distance $d_L\lesssim0.1$ Mpc for stellar-mass black holes) in which case LISA will start to be sensitive to them.  Alternatively, if the HCSC hosts IMBH ($M_{\bullet}\gtrsim100$ M$_\odot$) which experience gravitational bremsstrahlung themselves, detections with Advanced LIGO \citep{PhysRevD.102.062003} start to become feasible even at cosmological distances. The events considered here have bursts lasting between $\sim0.1\mu$s and $\sim120$s with median $45\mu$s. 
            
            Regarding EMRI candidates, the plot shows the frequencies and strains inferred from their final measured Keplerian orbital elements. We remind the reader that the inspiral is not directly measured due to the omission of relativistic effects here (Section~\ref{Sec:Methods}). As such, the points should be interpreted only as an approximate region in frequency-strain space where events are expected to lie. Indeed, previous research has shown how the inclusion of relativistic effects tends to increase the strain (and occurrence) of EMRI events \citep[i.e,][]{2018MNRAS.481.5445S, 2019PhRvD..99f3006S, 2021ApJ...907L..20T, 2022Natur.603..237S, 2024A&A...685A.123H}, as such many of the events indicated in the figure could in fact become observable, and with a greater rate. Beyond the omission of PN terms, we also investigate a relatively low-mass MBH, and heavier recoiling BHs will, not only result in increased strains, but also induce a greater merger rate (recall Equation~\ref{Eqn:Gamma_Fit} and Figure~\ref{Fig:Event_vs_gamma}).
    
            Lastly, head-on collisions between two white dwarfs may trigger a type Ia supernovae explosion \citep{2011NatCo...2..350H}. These bright events would be observable at cosmological distances and are characterised with a unique GW signal \citep{2012arXiv1211.4584K} making them particularly interesting to consider. Although no such events were recorded in the simulations, a more massive recoiling BH increases the likelihood given their more populated HCSC.
            
       \section{Discussion}
        \subsection{Observables forecasting} \label{Sec:Forecast}
            \subsubsection{The model}\label{Sec:TheModel}
            Using Equation~\ref{Eqn:Gamma_Fit} and its corresponding best-fit parameters, this section forecasts the event rate of mergers emerging from recoiling BHs who have been or in the process of being ejected from their host galaxy ($v_k\geq v_{\rm esc}\approx5\sigma$). Estimates ignore instrument limitations and directionality, resulting in upper limits. Using data from table \ref{Tab:EventQuant}, we assume a one-to-ten ratio between GW and TDE events. Note however that this ratio is sensitive to the assumed IMF and the extent of primordial mass segregation at the moment of MBH merger.
    
            The all-sky cumulative event rate up to some observed redshift, $z_{\rm obs}$, is calculated as,
            \begin{align}
                \dot{N}_{i}(<z_{\rm obs}) = \int_0^{z_{\rm obs}} \rho_i(z')\, \frac{dV_c(z')}{dz'}\, \frac{dz'}{1+z'}.
            \end{align}
            Here, $dV_{c}$ is the all-sky differential comoving volume element and $(1+z')^{-1}$ accounts for time dilation. Throughout this section, cosmological formulas are taken from \citet{1999astro.ph..5116H}, while cosmological parameters from \citet{2020A&A...641A...6P}. Calculations assume observations up to $z_{\rm obs}\leq3$.

            The term $\rho_i(z')$ is the volumetric event rate density at redshift $z'$. It folds in all HCSCs that exist at $z'$ by convolving their formation history with their age-dependent event rate such that,
            \begin{align}
            \rho_i(z) = \int_{z}^{z_{\max}} \mathcal{R}_m(z_{\rm form})\, \frac{dt}{dz_{\rm form}}\; \big\langle \Gamma_i(M_\bullet, v_k; t_{\rm age}) \big\rangle_{\theta\, \mid\, z_{\rm form}}\, dz_{\rm form}.
            \label{eq:rho}
            \end{align}
            Here $z_{\rm form}$ is the formation redshift of a HCSC, $t_{\rm age}$ is the age of the HCSC given as $t_{\rm age}\equiv t(z)-t(z_{\rm form})$ and $\theta=\{M_{\bullet}, v_k,\gamma\}$ the properties of a probabilistically sampled HCSC. The bracketed term is computed via a Monte-Carlo approach over $\theta$. We consider two models: $\gamma=1.75$ and $\gamma=1.0$.
            
            The integration over $z_{\rm form}$ is weighted by the cosmological history of IMBHB merger events via the term $\mathcal{R}_m(z)$. Values for $\mathcal{R}_m(z)$ at different redshifts are taken from cosmological simulations \citep{2025ApJ...991...58K}. Since \citet{2025ApJ...991...58K} find no mergers beyond $z>7$, we set $z_{\rm max}= 7$.  
            
            Integrating over $z_{\rm form}$ is essential since rates are sensitive to a HCSC's age. Depending on its age, the HCSC can either be in the initial burst phase, as seen here, the RR phase as described in KM08, or appear dormant. Additionally, event rates decay exponentially with time. For the burst phase this decay is expressed in Equation~\ref{Eqn:Gamma_Fit}. For RR, the decay is expected to diminish exponentially as $\exp{(-t_{\rm age}/\tau})$, where $\tau\approx (3.6GM_{\bullet}^2)/(v_k^2\langle m_*\rangle)$ (KM08). 

            Previous work has shown large scatter in the proportion of particles merging, with \citet{2012MNRAS.421.2737O} finding anywhere between $20\%-90\%$ and Hochart et al., {in prep.} anywhere between $25-50\%$. Given the wide range in values, it is assumed that $30\%$ of the cluster ends up being disrupted. This modifies the age at which HCSC appear dormant ($t_{\rm dor}$). When dormant, the HCSC contributes nothing to the merger rate. Test calculations showed that increasing this fraction leads to enhanced observables.
            
            Given the dependence of $t_{\rm age}$ on $\Gamma_i$, three temporal regimes are considered:
            \begin{enumerate}
                \item If $t_{\rm age}\leq t_{\rm dor}/3$, the rate follows from Equation~\ref{Eqn:Gamma_Fit}.
                \item If $t_{\rm dor}/3<t_{\rm age}(z)\leq t_{\rm RR}$, the rate follows from Equation~\ref{Eqn:RR_Rate_Body} and accounts for both the depleted HCSC population and change in characteristic decay time.
                \item If $t_{\rm RR}< t_{\rm age}\leq t_{\rm LB}(z)$: The system evolves over classic two-body relaxation mechanisms \citep[see][]{2012MNRAS.421.2737O}.
            \end{enumerate}
            The $1/3$ factor applied to $t_{\rm dor}$ in regime (1) when defining the boundary between the burst and RR regime originates from \citet{2018ApJ...853..141M}, who found that loss cone refuelling via the $\omega$-clustering is only efficient while $M_{\rm HCSC}(t)/M_{\rm HCSC,0}\gtrsim 1/3$. 
            
            If this factor was reduced, the burst phase would extend longer. While this can lead to temporarily enhanced $\dot{N}_i$, a greater fraction of HCSC will also be dormant. Test calculations showed that, overall, the factor had little influence since the burst phase only accounts for a small period of a HCSC's lifetime. Calculations do not consider the puffing up of the HCSC in time. This will artificially enhance rates. The assumption is done since the expansion of a HCSC takes $10^{8}$ yr to go from mpc to pc scales (Hochart et al., {in prep.}) and the rate is highly stochastic.
    
            The Monte Carlo procedure samples over $v_k$ and $M_{\bullet}$. The recoil kick probability distribution is taken from \citet{2012PhRvD..85h4015L}, who consider `wet' mergers and BHs with an isotropic spin distribution. A `wet' environment reduces the kick velocity since gravito-magnetic coupling between the BH and accretion disk will help align their spins \citep{2007ApJ...661L.147B, 2010MNRAS.402..682D}. Additionally, there can be both a `hot' (adiabatic index, $\Tilde{\gamma}=5/3$) and `cold' ($\Tilde{\gamma}=7/5$) merger. With a more pressure supported (rarefied) disk, `hot' mergers are less efficient at aligning spins resulting in kicks typically being a factor two greater than the `cold' regime. Little difference was seen between the cold and hot models since only the tail distribution of either model satisfy the regime we focus on ($v_{k}>v_{\rm esc}$). Following this, we only consider `hot' kicks.
            
            When sampling $M_{\bullet}$, galactic masses are drawn from the Press–Schechter formalism using results from \citet{2015MNRAS.450.4486F}. These galactic masses are then mapped to BH masses via the Häring–Rix relation \citep{2004ApJ...604L..89H}. Following our investigated mass range, we consider the regime: $10^{5}\leq M_{\bullet}$ [M$_\odot]\, \leq 4\times10^{5}$.
            
            We note that in their derivation of $\mathcal{R}_m(z)$ values, \citet{2025ApJ...991...58K} had assumed that every galaxy with stellar mass component $>10^{7}$ M$_\odot$ contain a NSC. Contrastingly, observations indicate an occupancy factor $\approx0.3-0.6$ \citep{2018ApJ...859...52O, 2019ApJ...878...18S}. Since binary IMBH mergers are expected to require NSCs \citep{2024ApJ...976...22K}, results here can lead to a factor two to three enhancement in forecasted rates.
            
            We validate our methodology by comparing sample calculations with those of \citet[hereafter SL12]{2012MNRAS.422.1933S}. These validating calculations considered larger BH masses, focused on a purely RR regime, fixed $\gamma=1$ and considered sources out to $z_{\rm obs}\lesssim0.5$. These choices mimic the procedure done in SL12 as best as possible, it wasn't clear how to tune other variables (i.e the IMF or galactic merger rates). 
            
            In the end, calculations here predict a factor $\approx 1.71$ greater than SL12. Differences are small, but arise from adopted cosmological models (with the one here lying on the upper end of predictions), our omission of instrument capabilities, us adopting a larger average stellar mass, as well as differences in the $M_{\bullet}-\sigma$ relation and assumed $f_{\rm dep}$ amongst other unknown differences. 
            
            \subsubsection{The rates}
            \begin{figure}
                \centering
                \includegraphics[width=.95\columnwidth]{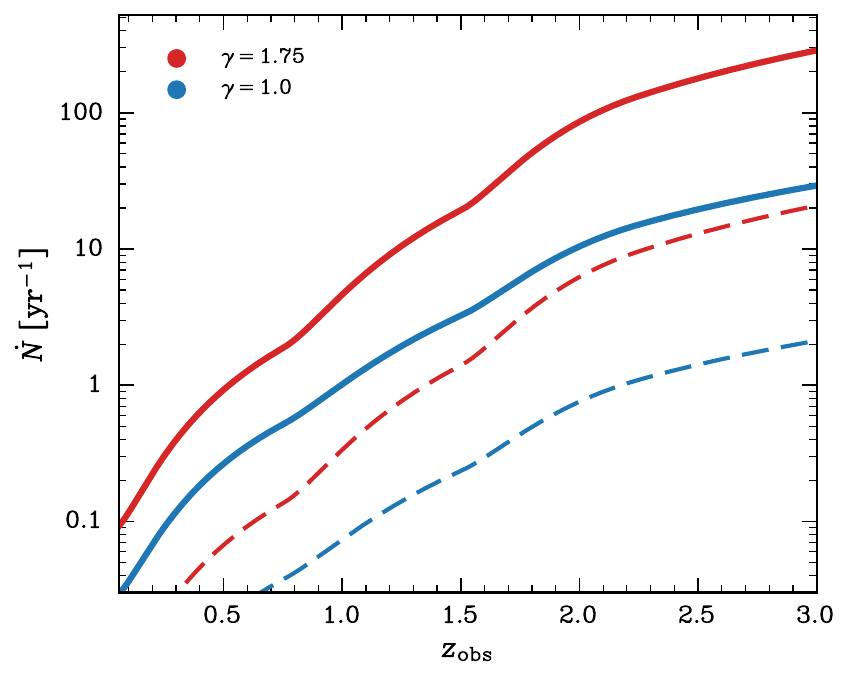}
                \caption{Cumulative merger rates up to $z<3$ for recoiling BHs ejected from their host galaxy. Solid lines denote the total number of observables, while dashed lines the number of GW events following our chosen IMF and without primordial mass segregation.}
                \label{Fig:Forecasting}
            \end{figure}
            When comparing $\gamma=1.75$ to $\gamma=1$, forecasting estimates predict a factor $\sim 10$ greater in event rates for the former. Although an order-of-magnitude larger, this remains smaller than that inferred by Figure~\ref{Fig:Event_vs_gamma}. This is because HCSC evolve quickly relative to cosmological time scales. As such, the prominence of dormant HCSC (more likely to occur for high $\gamma$), reduce the discrepancy. Even so, the results invoke a similar conclusion to that made in Figure~\ref{Fig:Event_vs_gamma}. Since rates are sensitive to $\gamma$, future observations of HCSC can reveal aspects of the inner parsec regions shortly before SMBHB or IMBHB coalescence.
            
            Considering only HCSC hosting a BH with mass between $10^{5}\leq M_{\bullet}$ [M$_\odot$] $\leq 4\times10^{5}$, we expect at most $\dot{N}\lesssim 290$ yr$^{-1}$ events up to $z_{\rm obs}\leq3$ when $\gamma=1.75$. Assuming our IMF holds throughout the cosmos, $\sim270$ of these would be TDEs. For $\gamma=1.0$, this reduces to $\dot{N}\lesssim 30$ yr $^{-1}$, with $\sim28$ being TDEs. We stress once more that the rate is sensitive to stochastic events such as the IMF and primordial mass segregation and here only quote values where $\alpha=-1.35$ and no primordial mass segregation is included.
            
            As mentioned in section \ref{Sec:TheModel}, here we assume all galaxies host a NSC. If the results from \citet{2024ApJ...976...22K} hold in that IMBHB mergers require a NSC, then predicted rates shrink by a factor $\sim0.3-0.6$. Even so, given the singular, low-mass IMBH observed to date \citep[$M_{\bullet}\approx142$ M$_{\odot}$,][] {2020PhRvL.125j1102A}, the notion that dozens of events involving these bodies are potentially observable with upcoming observatories is encouraging and may herald an exciting epoch in terms of understanding BH growth. 
            
            Regarding mass segregation, rates are found to be sensitive to the stellar mass and age distributions. The fiducial rates assume an average HCSC particle mass of $\langle m\rangle=1$ M$_\odot$. When increasing this to $\langle m\rangle = 2$ M$_\odot$, forecasting estimates give rates $\approx4.5$ times lower out to $z\leq3$. This difference is due to there being fewer bodies within the HCSC capable of being involved in a disruption event, and so systems are more often dormant.
            
            The large number of expected TDEs contradict with the lack of observations. Since compact objects segregate to the core easiest, the lack of observed offset TDEs may be partially due to primordial mass segregation. As such, the lack of observed offset TDEs would be compensated by a vast amount of EMRI's instead. Indeed, the inner $\sim10$ mpc is expected to be dominated by stellar mass black holes both for a NSC in steady-state \citep{2006ApJ...645.1152H, 2006ApJ...649...91F} and one whose recently had a IMBHB merger \citep{2026A&A...706A.354K}. Other possibilities can be due to the adopted recoil kick distribution overestimating high-velocity kicks or that the inner milliparsec regions of NCSs are significantly shallower than $\gamma=1.75$ at the moment of merger. The full scope of these effects lie beyond the present paper but represents a crucial direction for future work.
            
            Although a factor ten difference in rates emerges between $\gamma=1$ and $\gamma=1.75$, it is clear that constraining $\gamma$ isn't so straightforward since the problem is plagued by poorly constrained variables (i.e. IMF, stellar ages, mass segregation). This matter is complicated given the degenerate nature of the problem. Although frequent observations imply the combination of a steep $\gamma$ and large $\mathcal{R}_{m}(z')$, low rates can indicate either low $\mathcal{R}_{m}(z')$ or a shallow $\gamma$ at moment of merger. 
            
            Overall, future observations (or lack thereof) and ongoing developments in cosmology will further constrain the available parameter space. Alongside constraining $\gamma$ in the immediate vicinity of merged massive BHs, observations of HCSC could eventually constrain the IMF in galactic nuclei, the existence of IMBHB and IMBH seeds, the recoil kick distribution, the BH mass and spin distribution \citep{1998bhrs.conf...79R}.

            If the restriction of only focusing on ejected HCSC is removed, and rates consider systems kicked at $v_{k}>0.4v_{\rm esc}$ which corresponds roughly to the recoil kick needed to temporarily eject HCSCs from the bulge \citep{2008ApJ...678..780G}, a factor $\sim10-20$ is added to the fiducial rates. This enhancement increases for lower $\gamma$ and $M_{\bullet}$. The reduced kick means an increase to the relative bound HCSC mass (Equation~\ref{Eqn:MHCSC}). These won't necessarily all appear as offsets since the BH will oscillate back into the core over Gyr time scales
            
            Missions such as Chandra and eROSITA are able to detect the X-ray flares originating from TDEs at high redshift. This is exemplified with the $z\approx2.2$ transient event CDF-S XT1 observed by Chandra possibly being sourced by an IMBH-White Dwarf disruption \citep{2017MNRAS.467.4841B}. With a peak bolometric luminosity of $\sim10^{44}$ erg s$^{-1}$ lasting $\sim100$ days \citep{2021ARA&A..59...21G}, TDEs are particularly enticing for all-sky surveys such as eROSITA. In addition to CDF-S XT1, several offset TDE events have already been observed and may have an HCSC origin (i.e \citet{2016ApJ...821...25L, 2018ApJ...867...20C}) though this remains far under the predicted rates found here. The expected imminent launch of the Vera Rubin Observatory will help constrain how HCSC emerge, including those hosting a central IMBH. 
            
            Although results suggest a lack of events at high redshifts, the infrared capabilities of the James Webb Space Telescope will allow probing of these environments in the distant Universe via the thermal emission emerging from the resulting stellar debris. 
            
        \subsection{Little Red Dots}
            Little Red Dots (LRDs) are compact ($r < 100$ pc) systems at redshifts between $4\lesssim z\lesssim8$ \citep{2023ApJ...952..142F, 2024ApJ...963..129M, 2024ApJ...964...39G, 2024ApJ...977L..13B}. LRDs display active galactic nucleus (AGN)–like features such as UV and broad H$\alpha$ emission, yet often lack the X-ray and far-infrared signatures typically associated with accreting SMBHs \citep{2025ApJ...978...92L, 2024ApJ...969L..18A}. They also appear redder than expected for their redshift ($z>5$) and exhibit number densities exceeding those of UV-selected quasars. While some are isolated systems, others appear to be associated with galaxies, even appearing as offset nuclei \citep[see i.e;][]{2024A&A...691A..52K, 2025ApJ...983...60C, 2026Natur.653.1017J, 2025ApJ...992...71R, 2025A&A...698A.317M, 2026arXiv260106015Y}.
           
           Recent studies \citep{2025ApJ...984L..55B, 2025Univ...11..294P} suggest that TDEs can explain these observational signatures, with the former proposing that a subset of LRDs originate from runaway-collapsing clusters where the high densities allow for large TDE rates. For typical LRD masses and a required event rate of $\sim10^{-4}$ yr$^{-1}$ \citep{2025ApJ...984L..55B}, a young recoiling system with $M_{\bullet}=10^{7}\,{\rm M_{\odot}}$ and $v_k = 100$ km s$^{-1}$ can reproduce the observed luminosities for $\gamma \approx 0.9$ ($\gamma \approx 1.1$ for $M_{\bullet}=10^{6}\,{\rm M_{\odot}}$, see Figure~\ref{Fig:Event_vs_gamma}). Frequent TDEs during the early stages of a recoiling HCSC could thus mimic the radiative output of young stars and an accretion disk, producing the observed UV-bright continua and broad emission lines. This makes it tempting to identify a subset of LRDs as HCSCs. The possibility is made all the more enticing with some observations indicating that some LRDs have $f_b\equiv M_{\bullet}/M_{\rm *}\lesssim1$ \citep{2026Natur.653.1017J, 2026MNRAS.548f2109M}. This, however, only corresponds to a minority of LRDs \citep[see i.e;][]{2023ApJ...959...39H, 2024A&A...691A.145M, 2025ApJ...978...92L, 2025ApJ...983...60C, 2025arXiv251203239B}.
           
           While both these factors are encouraging, significant challenges remain. Newly formed HCSCs extend to $\mathcal{O}(10^{-2})$ pc scales (see Equation~\ref{Eqn:Rkick}) and require $0.1-1$ Gyrs to expand to parsec scales. While one LRD has had it's effective radius constrained to $R_e<30$ pc \citep{2024Natur.628...57F}, most extend to $\sim100$ pc \citep{2024ApJ...968...38K, 2024RNAAS...8..207G, 2025ApJ...991...37A}. Moreover, a recoiling origin faces spatial constraints: for a kick velocity of $v_k = 300$ km s$^{-1}$ and a travel time of 1 Gyr, the displaced HCSC would lie $\sim300$ kpc from its parent galaxy. That is roughly $40\%$ the distance to Messier 31 \citep{2012ApJ...745..156R}. As mentioned earlier, LRD's on the other hand, appear isolated. Combined with the fact that only a subset of LRDs exhibit emission variability characteristic of TDE flares \citep{2025ApJ...983L..26T, 2025ApJ...985..119Z} make us posit that HCSC can, at best, explain the faintest LRDs and only constitute a small subset of the population.

\section{Conclusions}
    Using direct $N$-body simulations, analysis of the burst of observables emanating from recoiling IMBHs ($M_{\bullet}=\{1,\ 4\}\times10^{5}$ M$_\odot$) are conducted. While we probe the IMBH mass regime, initial conditions are motivated by theory developed for SMBH mergers. Additionally, for completeness, the subsequent numerical results are extrapolated to the SMBH-mass regime. The main results are summarised below:
   \begin{enumerate}
       \item The instantaneous repopulation of the loss cone following a SMBHB or IMBHB coalescence, triggered by the recoil kick, produces a transient burst of gravitational-wave (GW) and tidal-disruption-event (TDE) activity. This burst phase lasts for several orbital periods, consistent with \citet{2018ApJ...853..141M} and \citet{2024ApJ...966L...4A}, and temporarily enhances the disruption rate from these systems.
       \item Assuming the inner $\sim10$ mpc of post-merger galactic nuclei follow a Bahcall–Wolf cusp, recoiling systems hosting a central IMBH and ejected from their host galaxy will yield up to $\lesssim 290$ offset TDEs and GWs per year up to redshifts $z<3$. If $\gamma=1$, forecasted rates decrease to $\lesssim30$ TDE and GW events per year.
      \item With event rates being sensitive to $\gamma$, future observations (or lack thereof) can provide insights on the density profile of the inner parsec of nuclear star clusters the moment of IMBHB merger. Observations will also inform models of galaxy–galaxy merger rates, the stellar initial mass function (IMF) within NSCs and the extent of mass segregation prior to massive BH mergers
   \end{enumerate}
    This study focuses on the burst phase of recoiling IMBH systems following dwarf-galaxy mergers. While several assumptions lead to optimistic forecasts, the resulting events should be detectable and may constitute smoking-gun evidence for IMBHB mergers, thereby constraining SMBH seed formation. Conversely, a null detection would imply either that IMBHB mergers are rare or that post-merger density profiles are shallow. The ratio between TDE-to-EMRI events will also provide insights on the mass function of the inner region within the NSC of galaxies post-merger.
   
   Amongst others, the forecasts presented here assume that the local $M_{\bullet}-r_{\bullet}$ relation holds at high redshift, consistent with recent observational evidence \citep{2026MNRAS.545f1979G, 2025ApJ...981...19L}. However, over-massive SMBHs have also been reported in the early Universe \citep[e.g.,][]{2024NatAs...8..126B, 2024ApJ...966L..30M}. Including this population of over-massive BHs would reduce the predicted TDE and GW rates found here. Indeed, if the $M_{\bullet}-r_{\bullet}$ relation is steeper than that found in \citet{2005SSRv..116..523F} ($\delta>0.59$ where $r_{\bullet}\propto M_{\bullet}^{\delta}$), most of the GW and TDE sources emanating from a given HCSC will orbit at relatively larger semi-major axis. With larger orbital periods, secular effects such as that driving the burst phase will also take longer. Meanwhile, the omission of PN terms underestimates the event rate. The effect is two fold; its inclusion reduces the precession timescale and introduces GW emission, which itself allows a tightening of orbits and enhancement of rates.

   Lastly, we do not perform analysis of the SMBH regime ($M_{\bullet}\gtrsim10^{6}$ M$_\odot$) as exploring this parameter space would require significantly greater computational resources. While improvements in numerical techniques have allowed researchers to probe NSCs in dwarf galaxies with mass resolutions on the order of $1$ M$_\odot$, \citep[i.e,][]{2024ApJ...976...22K, 2025MNRAS.537..956P}, extending such approaches to more massive systems remains computationally prohibitive. Consequently, it is difficult to motivate any extrapolation of our results to SMBH masses. A treatment of this regime is left for future work.
    
\section{Energy consumption}
    The $25$ simulations conducted during this investigation had a total wall-clock time of $43$ days, with each run using $5$ cores. In total, the CPU time for all simulations was $215$ days or $5160$ hours . The relevant node uses Intel Xeon Gold 5220R processors, which uses $150$ W. The total energy consumed is thus $E\approx 5160\times150\times\frac{1}{24}=32.250$ kWh.
    
    In 2025, the Netherlands had a CO$_2$ equivalent of $0.388$ kg kWh$^{-1}$, of which $28\%$ of the energy produced was renewable\footnote{https://www.nowtricity.com/country/netherlands/}. Factoring this green energy, the kWH to CO$_2$ equivalent emission is $0.27936$ kg kWh$^{-1}$. In turn, the total equivalent CO$_2$ emission is $E_{\rm CO_2}\approx (91.875\, {\rm kWh})\times(0.27936 {\rm\, kg\, kWh^{-1}})\approx9.0$ kg. That is roughly the amount of CO2 absorbed by a single adult female ginkgo tree after 2.5 months \citep{2024EgJB...64..258S}.
 
\begin{acknowledgements}
    We thank the referees for being very critical in our earlier versions. Their feedback greatly strenghtened the quality and validity of the paper. It is also a pleasure to thank Daming Yang and Mark Gieles for insightful discussions. In addition, we would also like to thank Konstantinos Kritos for providing results for the IMBH-IMBH merger rates in our mass range. Analysis was made using the open-source \texttt{Python} packages \texttt{NumPy} \citep{2020Natur.585..357H} and \texttt{Matplotlib} \citep{2007CSE.....9...90H}.
\end{acknowledgements}

\bibliographystyle{aa}
\bibliography{references.bib}

\begin{appendix}
    \section{Deriving the Event Rate}
    The simulations here are integrated for $t=0.1$ Myr. This is much shorter than RR timescales of $t_{\rm RR}\approx10^{3}$ Myr (c.f Equation~\ref{Eqn:Gamma_Relation}). Given this, RR is not the correct mechanism to consider for event rates. 
    
    To illustrate this, consider the analytical event rate predicted using RR \citep{1996NewA....1..149R, 1998MNRAS.299.1231R, 2006ApJ...645.1152H},
    \begin{equation}
        \Gamma \approx C_{\rm RR}(\gamma)\frac{\ln \Lambda}{\ln(r_k/R_{\rm tide})}\left(\frac{v_k}{r_k}\right)f_b. \label{Eqn:RR_Rate}
    \end{equation}
    Here $C_{\rm RR}$ is some numerical constant, dependent on the environment while $\ln \Lambda$ is the Coulomb logarithm. The other variables have already been introduced in the main paper, with $r_k$ and $R_{\rm tide}$ following Equation~\ref{Eqn:Rkick} and Equation~\ref{Eqn:Rtide} respectively. Using $f_b\equiv M_{\rm HCSC}/M_{\bullet}$ and Equation~\ref{Eqn:MHCSC}, one gets back,
    \begin{equation}
        \Gamma\propto M_{\bullet}^{2-\gamma}\,v_k^{2\gamma-3},
    \end{equation} 
    a relation not observed here as made most apparent by the inverse $v_k$ relation in results (recall Figure~\ref{Fig:ALL_Events}). Instead, events here are predominantly sourced from the instantaneous refilling of the loss cone by the kick (see \citet{2011MNRAS.412...75S}) and the resulting eccentric, $\omega$-clustered population which emerges. 
    
\begin{figure}
        \centering
        \includegraphics[width=\columnwidth]{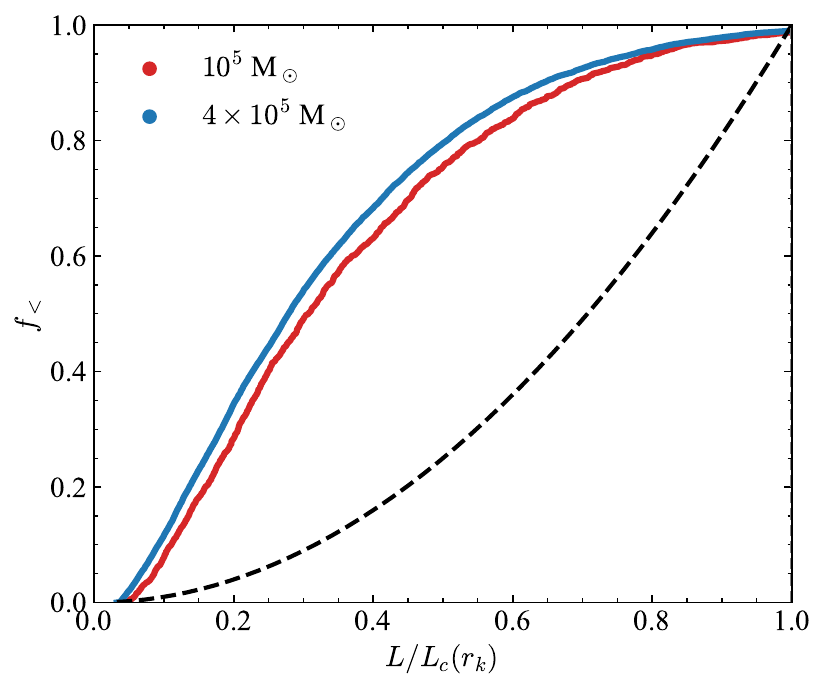}
        \caption{Cumulative distribution function of the angular momentum normalised by the circular value at $r_k$ for particles at the moment of recoil ($v_k=300$ km s$^{-1}$). The dashed line represents an isotropic environment.}
        \label{Fig:AngMom}
    \end{figure}

    At the moment of kick, particles all experience a velocity shift $-v_k$ relative to the BH, resulting in a change in angular momentum:
    \begin{equation}
        L' = |r\times (v-v_{k})| \approx rv_k\sin\phi(\omega),
    \end{equation}
    where $\phi$ is the angle between the kick direction and the orbital velocity at periapse. The effect of the recoil drives many of the bound particles to occupy lower angular momentum orbits. This is seen in Figure~\ref{Fig:AngMom}, since the cumulative distribution of the bound particles' angular momentum is shifted toward lower $L$ relative to the isotropic case.
    Before proceeding, we make two approximations. First, since orbits are near Keplerian, we work with orbit-averaged quantities, allowing us to convert the instantaneous position of a particle relative to the central MBH to its semi-major axis ($r\xrightarrow[]{}a$). Second, we assume HCSC have a non-spherical potential.
    
    To support this, we make use of a particle mass-weighted tensor,
    \begin{equation}
        S_{ij}\equiv\frac{\sum_{k}m_k x_{k,i}x_{k,j}}{\sum_k m_k},
    \end{equation}
    where $x_{k}$ is the position of particle $k$ relative to the SMBH and $m_k$ the particle's mass. The eigenvalues of this tensor, $a^2\geq b^2\geq c^2$ correspond to the three principal axis lengths. Computing their ratios will describe how spherical ($a\approx b\approx c$) the system is. 

\begin{figure}
        \centering
        \includegraphics[width=\columnwidth]{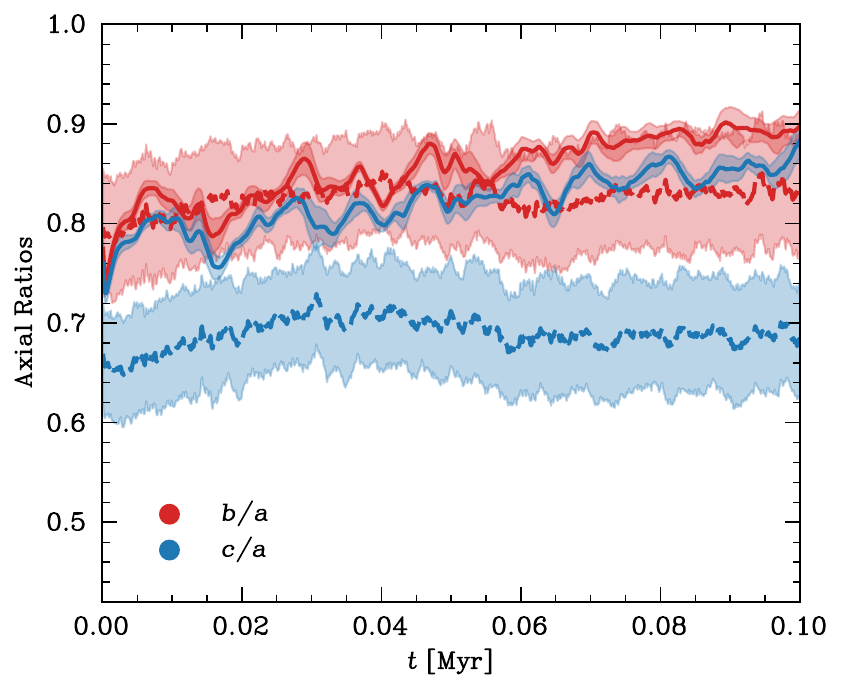}
        \caption{The ratio between the three semi-axis of the HCSC when $v_{k}=600$ km s$^{-1}$ and $M_{\bullet}=10^{5}$ M$_\odot$ (dashed) and $v_k=300$ km s$^{-1}$ and $M_{\bullet}=4\times10^{5}$ M$_\odot$ (solid).}
        \label{fig:Ellipsoid}
    \end{figure}

    Figure~\ref{fig:Ellipsoid} shows that $b/a\approx 0.8-0.9$ throughout the simulation for either configurations shown. Meanwhile, $c/a$ spans a wider range of values. Since both deviate from unity, particularly at early times and for $c/a$, the system is seen to be mildly triaxial, consistent with \citet{2023ApJ...958..137A}. In the case of $v_k=300$ km s$^{-1}$ and $M_{\bullet}=4\times10^{5}$ M$_\odot$, $b/c\approx 1$ meaning the system is near axisymmetric.
    
    Given this, section 3 of \citet{2013CQGra..30x4005M} tells us that the geometric fraction of particles within the BH loss cone in a non-spherical potential goes as,
    \begin{equation}
        P_{\rm LC}(a) \sim \ell\ell_{\rm lc}.
    \end{equation}
    Here $\ell\equiv L/L_{\rm LC}$ and $\ell_{\rm lc}\equiv L_{\rm min}/L_{\rm lc}$. Near the MBH $\ell\sim\sqrt{\epsilon}\ell_{\rm lc}$ where $\epsilon$ is some factor dependent on ratio between an orbits semi-axis \citep{2013CQGra..30x4005M}. As a result, one can write,
    \begin{align}
        P_{\rm LC}(a)&\sim \sqrt{\epsilon}\frac{L_{\rm min}}{L_{\rm lc}} \\
        P_{\rm LC}(a)&\sim \sqrt{\frac{1}{2}(1-q)}\frac{\sqrt{2GM_{\bullet} R_{\rm tide}}}{\langle m_*\rangle v_ka},
    \end{align}
    where $q\equiv c/a$.

    A 1D geometry characterised by the $(v_k a)^{-1}$ term is used because the system hasn't yet relaxed (isotropised) from the kick with the system being highly correlated since all particles have the same velocity translation. If the system was isotropised, $P_{\rm LC}\propto (L_{\rm min}/av_k)^{2}$ \citep{2013CQGra..30x4005M}. In this case, however, with a nonspherical potential, asymmetric torques act to change a particles angular momentum, resulting in a greater loss-cone repopulation \citep[see i.e,][]{2001ApJ...549..192P, 2004ApJ...606..788M, 2006astro.ph..1520H, 2006ApJ...642L..21B, 2015ApJ...810...49V, 2017MNRAS.464.2301G}.
    
    For what is to follow, ideally one should consider the effective velocity, $v_{\rm eff}=\left({v_k^2+v_{\rm orb}^2}\right)^{1/2}$ since $v_k\approx v_{\rm orb}$ (see table \ref{Tab:AppKick}). However, to simplify the algebra we set $v_{\rm eff}\approx \sqrt{2}|v_{k}|$. In all cases, a factor $v_k^{-1}$ term survives so the intuition remains the same. Note also the use of $R_{\rm tide}\propto M_{\bullet}^{1/3}$ rather than the innermost stable circular orbit, $R_{\rm ISCO}\propto M_{\bullet}$. For larger BHs ($M_{\bullet}\gtrsim 10^{8}$ M$_\odot$), $R_{\rm tide}<R_{\rm ISCO}$ and so the assumption falters. Overall, if $R_{\rm ISCO}$ is considered, an additional $M_{\bullet}^{1/3}$ dependence appears in the final equation (c.f Equation~\ref{Eqn:FinalEqn}). 
    
    Due to the efficient merging of the binary BHs when $a\lesssim a_{\rm GW}$, it was initially assumed that no particles have periastron $r_p<a_{\rm GW}$. This restriction of $(E,L)$-space translates into an initial maximum eccentricity of,
    \begin{equation}
        a_{\rm GW} = a(1-e_{\rm max}),
    \end{equation}
    \begin{equation}
        \therefore e_{\rm max} = 1 - \frac{a_{\rm GW}}{a}.
    \end{equation}
    Assuming a thermal distribution, $f(e)=2e\, de$ \citep{1919MNRAS..79..408J}, the factor which accounts for the initially removed population who had $r_p<a_{\rm GW}$ is,
    \begin{equation}
        S_{r_p} = \int_{0}^{e_{\rm max}} 2e\, de = \left(1-\frac{a_{\rm GW}}{a}\right)^{2}. \\
    \end{equation}
    This factor reduces the proportion of events being sourced at $a\approx a_{\rm GW}$, pushing the dominant $a$ of disrupted particles to outer regions. $S_{r_p}$ also motivates our ignoring of the effective velocity, since if this wasn't considered and $r_p<a_{\rm GW}$ remained, low-$a$ sources orbiting with $v_{\rm orb}\gg v_{k}$ will play a larger role. 
    
    \begin{table}
        \centering
        \caption{Median orbital velocities of particles who end up merging with the BH the moment of kick. Row 1: $v_k=300$ km s$^{-1}$. Row 2: $v_{k}=600$ km s$^{-1}$.}
        \begin{tabular}{c c} \hline
             $M_{\bullet}=10^{5}$ M$_\odot$ & $M_{\bullet} =4\times10^{5}$ M$_\odot$\\ \hline
              $487^{+425}_{-233}$ km s$^{-1}$ & $621^{+590}_{-306}$ km s$^{-1}$   \\
              $657^{+374}_{-254}$ km s$^{-1}$ & $815^{+559}_{-396}$ km s$^{-1}$ \\
              \hline
        \end{tabular}
        \label{Tab:AppKick}
    \end{table}
    Overall, the event rate within the range $r\in[a,\, a+da]$ is defined as,
    \begin{equation}
        \Gamma_0 = \int_{a_{\rm GW}}^{r_k} \frac{N(a)}{P_{\rm orb}(a)}\times P_{\rm LC}(a) \times S_{r_p}\, da. \label{Eqn:EventRate_Theory}
    \end{equation}
    Here, the first term represents the number of stars, $N(a)$, approaching periapsis per unit time. This unit of time is measured by the orbital period, $P_{\rm orb}(a)$, and taken from Kepler's third law \citep{1609anov.book.....K}. The second term represents the fraction of stars whose angular momentum lies within the IMBH loss cone, and the third term, as mentioned prior, reduces the weight of events from $a\sim a_{\rm GW}$ stars simply because few remain given the eccentricity distribution. 
    
    The integrand is considered over the range of $L(a_{\rm GW})$ and $L_{\rm max}$ such that $a\in[a_{\rm GW},\, r_{k}]$. The lower bound does not consider $L_{\rm min}$ since, once more, no stars lie within $r_{ij}\leq a_{\rm GW}$ due to the binary scouring that population and relaxation processes not having enough time to repopulate it. 
    
    Focusing on the first term. For a power-law cusp, $\rho\propto r^{-\gamma}$.  The number of stars within a shell is:
    \begin{equation}
        N(r) = 4\pi r^2 \rho(r)\, dr,
    \end{equation}
    where for a power-law cusp,
    \begin{equation}
        \rho(r)=\rho_0\left(\frac{r}{r_0}\right)^{-\gamma}.
    \end{equation}
    Given this, the number of particles within a shell can be written as,
    \begin{equation}
        N(r) = 4\pi\rho_0 r_0^\gamma\,  r^{2-\gamma}\, dr.
    \end{equation}
    The normalisation factor $\rho_0r_0^\gamma$ acts as a normalisation constant since the recoiled HCSC does not have an arbitrary density profile at initialisation. Instead, it was initially embedded within an NSC constructed with the $M_{\bullet}-\sigma$ relation in mind. The normalisation factor $\rho_0r_0^\gamma$ can be found by noting that the enclosed mass at some radius is,
    \begin{equation}
        M(<r)=\frac{4\pi}{3-\gamma}\left(\rho_0r_0^\gamma r^{3-\gamma}\right),
    \end{equation}
    such that upon rearrangement,
    \begin{equation}
        \rho_0 r_0^\gamma = \frac{3-\gamma}{4\pi}M(<r) r^{\gamma-3}.
    \end{equation}
    Setting $r=r_k$ such that $M(<r_k)=M_{\rm HCSC}$, Equation~\ref{Eqn:MHCSC} tells us that,
    \begin{equation}
        \rho_0 r_0^\gamma = \frac{3-\gamma}{4\pi}\left(M_{\bullet}\, F_1(\gamma)\,  (8r_{\bullet})^{\gamma-3}\right). \label{Eqn:rho0_r0}
    \end{equation}
    This will be useful later.
    
    Using Kepler's third law, the first term can be expressed as,
    \begin{equation}
        \frac{N(a)\, da}{P_{\rm orb}(a)} \propto \frac{(\rho_0 r_0^\gamma) a^{2-\gamma}\, da}{\sqrt{a^3/M_{\bullet}}}.
    \end{equation}
    From here on out, only proportionalities will be considered to reduce clutter. An explicit expression will also be given at the end for generalisability. 
    
\begin{figure}
        \centering
        \includegraphics[width=\columnwidth]{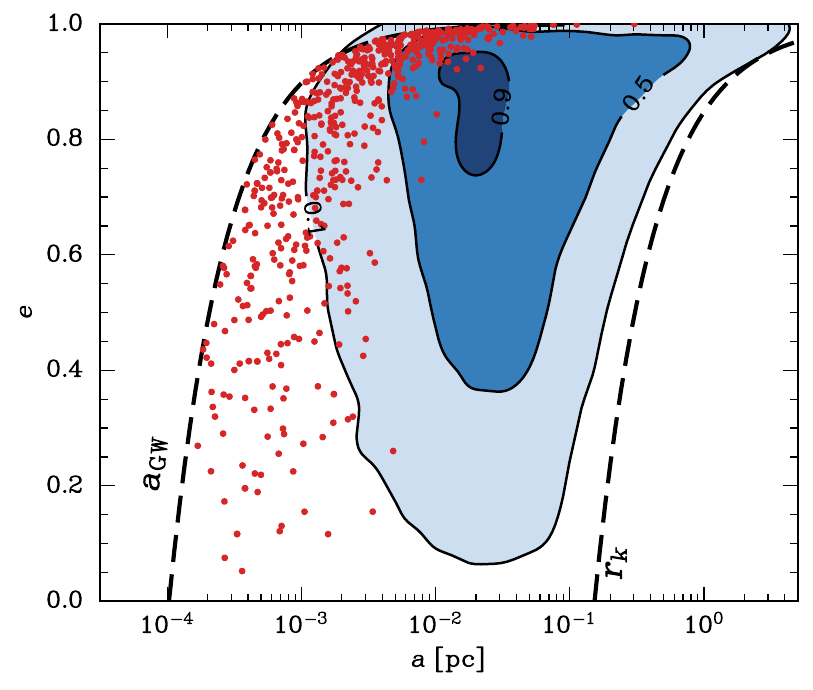}
        \caption{The orbital parameters of bodies during all $M_{\bullet}=4\times10^{5}$ M$_\odot$, $v_{k}=300$ km s$^{-1}$ runs. The heatmap shows the distribution of semi-major axis and eccentricity of bound particles the moment a GW recoil kick has been applied on the IMBH while the scattered dots show the initial parameters for bodies who eventually merge with the central IMBH. The two dashed lines signify the HCSC radius, $r_k$ (Equation~\ref{Eqn:Rkick}), and the region where GW emission dominates the preceding IMBH binary's evolution, $a_{\rm GW}$, multiplied by a factor $(1-e)^{-1}$.}
        \label{fig:App1}
    \end{figure}

    Since all three terms within Equation~\ref{Eqn:EventRate_Theory} is now known, Equation~\ref{Eqn:EventRate_Theory} can be solved. Integrating the function gives,
    \begin{align}
        \Gamma_0 &\propto \int_{a_{\rm GW}}^{r_k}\left(\frac{\rho_0 r_0^\gamma\, a^{2-\gamma}}{\sqrt{a^3/M_{\bullet}}}\right)\times\left(\frac{\sqrt{2GM_{\bullet}R_{\rm tide}}}{\langle m_*\rangle v_k a}\right)\times\left(1-\frac{a_{\mathrm GW}}{a}\right)^{2}\, da, \\
        \Gamma_0 &\propto \rho_0 r_0^\gamma \frac{M_{\bullet}\sqrt{R_{\rm tide}}}{\langle m_*\rangle v_k} \int_{a_{\rm GW}}^{r_k} a^{-(\gamma+1/2)}\times\left(1-\frac{a_{\rm GW}}{a}\right)^2\, da.
    \end{align}
    Setting $x\equiv a/a_{\rm GW}$ such that $da=a_{\rm GW}\, dx$ and the upper and lower limits become $r_{k}/a_{\rm GW}$ and $1$ respectively,
    \begin{equation}
        \Gamma_0 \propto \rho_0 r_0^\gamma \frac{M_{\bullet}\sqrt{R_{\rm tide}}}{\langle m_*\rangle v_k} a_{\rm GW}^{-(\gamma-1/2)}\int_{1}^{r_k/a_{\rm GW}} x^{-(\gamma+1/2)}\left(1-\frac{1}{x}\right)^2\, dx
    \end{equation}
    Looking at the integrand only:
    \begin{equation}
        \mathcal{I} = \left(-\frac{2\left(\left(2\gamma+3\right)x\left(\left(2\gamma+1\right)x-4\gamma+2\right)+4\gamma^2-1\right)}{\left(2\gamma-1\right)\left(2\gamma+1\right)\left(2\gamma+3\right)x^{(2\gamma+3)/2}} + C\right) \Bigg|^{r_k/a_{\rm GW}}_{1}
    \end{equation}
    This is a converging integrand since for most of the relevant parameter space $r_k\gg a_{\rm GW}$. To illustrate this point, consider the case where $r_k\approx a_{\rm GW}$. Equating Equation~\ref{Eqn:aGW} with Equation~\ref{Eqn:Rkick} with help of Equation~\ref{Eqn:FerrareseFord} and \ref{Eqn:rinfl}, one can derive the requirement $v_k\approx200 \sigma$. For a Milky Way-like system, this requires kicks $v_k>2\times10^{4}$ km s$^{-1}$. For a dwarf galaxy with IMBH of mass $M_{\bullet}=10^{5}$ M$_\odot$, $v_k\gtrsim 9\times10^{3}$ km s$^{-1}$. This lower-bound being the dominant source location is also partially motivated by Figure~\ref{fig:App1}. As a result, its safe to assume the lower bound dominates the integral. 
    
    Considering this, define,
    \begin{equation}
        A \equiv \frac{2\left(\left(2\gamma+3\right)\left(\left(2\gamma+1\right)-4\gamma+2\right)+4\gamma^2-1\right)}{\left(2\gamma-1\right)\left(2\gamma+1\right)\left(2\gamma+3\right)}, \label{eqn:Acoeff}
    \end{equation}
    such that,
    \begin{equation}
        \Gamma_0 \propto A\, \rho_0 r_0^\gamma \frac{M_{\bullet}\sqrt{R_{\rm tide}}}{\langle m_*\rangle v_k} a_{\rm GW}^{-(\gamma-1/2)}.
    \end{equation}
    Using Equation~\ref{Eqn:rho0_r0} for $\rho_0r_0^\gamma$, Equation~\ref{Eqn:Rtide} for $R_{\rm tide}$ and Equation~\ref{Eqn:aGW} for $a_{\rm GW}$,
    \begin{equation}
        \Gamma_0 \propto A\cdot \left((3-\gamma)\cdot M_{\bullet}r_{\bullet}^{\gamma-3}\right)\cdot \frac{M_{\bullet}\left(M_{\bullet}^{1/6}\right)}{\langle m_*\rangle v_k}\cdot\left(r_{\bullet}^{-(\gamma-1/2)}\right).
    \end{equation}
    Using the \citet{2005SSRv..116..523F} relation, $r_{\bullet}\propto M_{\bullet}^{\delta}$ where $\delta\approx0.59$ and setting $\gamma = 1.75$ as per the simulations here,
    \begin{equation}
        \Gamma_0 \propto A\, \left(3-\gamma\right) M_{\bullet}^{0.7} v_k^{-1}.\label{Eqn:Gamma0} 
    \end{equation}
    
    To make the expression complete, analogous to KM08, the event rate is predicted to drop off as $\Gamma(t)=\Gamma_0e^{-t/\tau}$ where $\tau$ is some damping factor (more on this shortly). Taking the limits for intuition,
    \begin{align}
        t&=t_{\rm sim}\xrightarrow[]{} 0,& \exp\left({-t/\tau}\right) \xrightarrow[]{} 1 \\
        t&=t_{\rm sim}\xrightarrow[]{}\infty, &\exp\left({-t/\tau}\right)\xrightarrow[]{}0
    \end{align}
    which agrees with our results, since while $v_{k}=600$ km s$^{-1}$, $M_{\bullet}=10^{5}$ M$_\odot$ has some damping by $t_{\rm sim}\approx\tau= 50$ kyr (see Figure~\ref{Fig:ALL_Events}), the other configurations where $\tau>t_{\rm sim}$ preserve a linear trend in event rates.
    
    Choosing $\tau$ is non-trivial, and requires understanding that immediately after the recoil kick, particles acquire a coherent apsidal alignment \citep{2018ApJ...853..141M}. For an in-plane kick along $+\hat{x}$, the mean eccentricity vector, which points towards the periapsis, goes as \citep{2021ApJ...921L..12A},
    \begin{equation}
        \langle e\rangle = \frac{3}{2}\frac{v_{k}}{v_{\rm orb}}\hat{y}.
    \end{equation}
    That is, the alignment is perpendicular to the kick in the orbital plane, and a tight anisotropy forms in $\omega$. This alignment is also shown in figure 2 and figure 8 of \citet{2023ApJ...958..137A} and occurs while $v_{\rm orb}\sim v_{k}$, as the case here (recall Table 1). 
    
    This clustering is important to consider since \citet{2018ApJ...853..141M} and \citet{2021ApJ...921L..12A} find that the torques induced by the clustered population on non-clustered neighbours can allow refilling of the loss-cone to persist up to $>10^{4}$ orbital periods, extending the burst phase.  While the overall duration of this phase remains unknown, \citet{2018ApJ...853..141M} find that the mechanism remains efficient as long as at least one third of the initial cluster mass is retained.
    
    The extent to which the mechanism is more efficient in refilling the loss cone can also be seen by comparing its timescale with the RR timescales. The relevant timescales go as \citep{1996NewA....1..149R},
    \begin{align}
        t_{\omega} &\approx \frac{M_{\bullet}}{M_{\rm HCSC}} P_{\rm orb}, \\
        t_{\rm RR} &\approx \frac{M_{\bullet}}{\langle m_*\rangle} P_{\rm orb}.
    \end{align}
    Since $M_{\rm HCSC}\gg\langle m_*\rangle$, $t_{\omega}$ acts on much shorter timescales and is much more efficient in refilling the loss-cone. As a result, since KM08 consider only RR effects when refilling the loss-cone, they underestimate the rates present within recoiled systems.
    
    The mechanism behind repopulating the loss cone is as follows: if a particle precesses ahead of the clustered population, the torque exerted by the clustered population onto it increases the particle's eccentricity while leaving the orbital energy unchanged. As a result, the angular momentum, $L$, decreases. This effect also causes it's precession rate to increase since for the same $a$, a higher $e$ precesses quicker \citep{2018ApJ...853..141M, 2024ApJ...966L...4A}. If, instead, the particle is lagging behind, the opposite effect occurs. The oscillation in a particle's precession rate and eccentricity induces grazing orbits and repopulates the loss cone.
    
    \citet{2018ApJ...853..141M} find that the mechanism acts periodically, with a characteristic timescale
    \begin{equation}
        \tau_{\rm osc}=\frac{2\pi}{\Delta \omega} =(e\phi_{\rm disk})^{3/2}t_{\omega}.
    \end{equation}
     Here, $\phi_{\rm disk}$ is the angular spread of eccentricity vectors within the clustered population. Hereafter $\tau_{\rm osc}\approx t_{\omega}$. This is the term which appears in the exponential and accounts for the decaying HCSC population. 
     
     Temporarily considering results here, plugging in $M_{\bullet}=10^{5}$ M$_\odot$, $v_k=600$ km s${^-1}$ yields $\tau_{\rm osc}\approx t_{\omega}\approx45$ kyr. When compared to Figure~\ref{Fig:ALL_Events} this passes the eye test given the clear deceleration in rates occuring at around that time. For $v_{k}=300$ km s$^{-1}$, $t_{\omega}\approx 65$ kyr, and the slight dampening also appears within the figure.
    
    As before, looking at proportionalities only and substituting Equation~\ref{Eqn:MHCSC} to find $t_{\omega}$, 
    \begin{equation}
        t_{\omega} \propto \frac{3-\gamma}{\rho_0 r_0^\gamma r^{3-\gamma}}\cdot P_{\rm orb}(a)M_{\bullet}.
    \end{equation}
    Since most of the sources are sourced at $\sim a_{\rm GW}$, we consider this region of the enclosed mass since it is their precession time scale we care for. That is, $r= \zeta a_{\rm GW}$ for some $\zeta$ such that,
    \begin{equation}
        t_{\omega} \propto \frac{3-\gamma}{\rho_0 r_0^\gamma a_{\rm GW}^{3-\gamma}}\cdot P_{\rm orb}(a)M_{\bullet},
    \end{equation}
    which, when substituting Equation~\ref{Eqn:aGW} and Kepler's third law,
    \begin{align}
        t_{\omega} &\propto \left(M_{\bullet}r_{\bullet}^{3-\gamma}\, r_{\bullet}^{\gamma-3}\right)^{-1}\cdot \sqrt{\frac{r_{\bullet}^3}{M_{\bullet}}}M_{\bullet} \\
        t_{\omega}&\propto M_{\bullet}^{0.385}. \label{Eqn:t_omega}
    \end{align}
    The last line used the relation found in \citet{2005SSRv..116..523F}.

    Finally, one should consider that the mechanism only acts while a particle interacts with the clump. As of now, Equation~\ref{Eqn:Gamma0} assumed the supply rate was constant. To account for this `window of repopulating the loss cone', a factor $f_{\rm win}\equiv(\tau_{\rm int}/\tau_{\rm osc})$ has to be introduced, where $\tau_{\rm int}$ accounts for the interaction time between the clump and some particle $i$. 
    
    The interaction is only efficient when the particle and mass clump are within some angle, $\varphi_{\rm max}$ of one another. The resulting interaction time is then,
    \begin{equation}
        \tau_{\rm int}\equiv\frac{2\varphi_{\rm max}}{|\Delta\dot{\varphi}|},
    \end{equation}
    The denominator represents the relative angular velocity between the two, and follows,
     \begin{equation}
         \Delta\dot{\varphi} =|n_i-n_C|=\Big|\frac{2\pi}{P_{{\rm orb},\, i}(a_i)}-\frac{2\pi}{P_{{\rm orb},\, c}(a_c)}\Big|.
    \end{equation}
    Here, $P_{{\rm orb},\, c}(a_c)$ is the orbital period of the clump at semi-major axis $a_c$ relative to the central (recoiling) BH.
    
    For the numerator, we assume that $\varphi_{\rm max}$ depends on the Hill-sphere of the clustered mass. This is motivated by the fact that it needs to perturb the particle to exchange angular momentum with it and it allows us to scale $\varphi_{\rm max}$ in relation to the Hill radius such that,
    \begin{equation}
         \varphi_{\rm max} \approx k\frac{R_{\rm Hill}}{|a_i-a_c|} = k \frac{a_c}{|a_i-a_c|}\left(\frac{M_{c}}{M_{\bullet}}\right)^{1/3}.
    \end{equation}
    Here, $k$ is some constant and $M_c$ is the clump mass. The relative semi-major axis differ by a numerical factor and so are omitted in later steps. Using Kepler's third law again, it can be shown that ,
    \begin{equation}
        \tau_{\rm int} \approx \frac{k}{\pi\sqrt{GM_{\bullet}}} \left(\frac{M_c}{M_{\bullet}}\right)^{1/3} \Big| \left(\frac{\beta^{3/2}a_i^{3/2}}{\beta^{3/2}-1}\right)\Big|,
    \end{equation}
    where $\beta\equiv a_c/a_i$. Substituting the, now familiar, $M_{\bullet}-r_{\bullet}$ relation which set the initial conditions and accounting for the fact that events are sourced at $a\approx\zeta a_{\rm GW}\propto r_{\bullet}$ such that $a_i\propto r_{\bullet}$,
    \begin{align}
        \tau_{\rm int} &\propto \frac{1}{\sqrt{GM_{\bullet}}}\left(\frac{M_c}{M_{\bullet}}\right)^{1/3} r_{\bullet}^{3/2}, \\
        \tau_{\rm int} &\propto M_c^{1/3}M_{\bullet}^{0.05} \label{Eqn:tau_int}.
    \end{align}
    Finally, using this relation along with Equation~\ref{Eqn:t_omega}, then the total event rate when assuming $\gamma=1.75$ can be written as,
    \begin{align}
        \Gamma(t) &\equiv \Gamma_0\,\frac{\tau_{\rm int}}{\tau_{\omega}}\,e^{-t/\tau_{\omega}}, \\
        \Gamma(t) &\propto A\, M_{\bullet}^{0.365}v_k^{-1}.
    \end{align}
    The full equation is given as,
    \begin{align}
        \Gamma &\approx \frac{3-\gamma}{\sqrt{2}\pi\langle m_*\rangle} \sqrt{\frac{1-q}{2}} GM_{\bullet}^2\left(R_*^3\eta^2 \frac{M_{\bullet}}{\langle m_*\rangle}\right)^{1/6}\left(F_1(\gamma)\left(8 r_{\bullet}\right)^{\gamma-3}\right)\, v_k^{-1} \notag \\
          &\hspace*{1.5em} \times a_{\rm GW}^{-(\gamma-1/2)}\left(\frac{2\left(\left(2\gamma+3\right)\left(\left(2\gamma+1\right)-4\gamma+2\right)+4\gamma^2-1\right)}{\left(2\gamma-1\right)\left(2\gamma+1\right)\left(2\gamma+3\right)}\right) \notag \\
         &\hspace*{1.5em} \times\left(\frac{2\pi\left(2\times10^{-4}\, e\phi_{\rm disk}\right)^{3/2}r_{\bullet}^{3/2}}{\sqrt{GM_{\bullet}}}\cdot \left(F_1(\gamma)\left(\frac{4\times10^{4}}{\zeta}\right)^{\gamma-3}\right)\right)^{-1} \notag \\
         &\hspace*{11em} \times\frac{k M_c^{1/3}}{\pi M_{\bullet}^{1/3}\sqrt{GM_{\bullet}}}\cdot \left(\frac{\beta^{3/2}a_i^{3/2}}{\beta^{3/2}-1}\right) \label{Eqn:FinalEqn}
    \end{align}
    Here $\zeta$ is defined such that sources are predominantly sourced at $\zeta a_{\rm GW}$ and $\beta\equiv a_{c}/a_i$. The equation can be further simplified but all terms remain (namely $F_1(\gamma)$) for transparency.
    \begin{align}
        \Gamma &\approx \frac{2.5\times10^{5}}{\pi^3}\frac{G\sqrt{R_*}\eta^{1/3}}{\langle m_*\rangle^{7/6}v_k} \frac{kM_c^{1/3}(8r_{\bullet})^{\gamma-3}M_{\bullet}^{11/6}}{a_{\rm GW}^{\gamma-1/2}(e\phi_{\rm disk}r_{\bullet})^{3/2}}\frac{(\beta a_i)^{3/2}}{\beta^{3/2}-1}\notag \\
        &\hspace{1.5em}\times\frac{(3-\gamma)\left(\left(2\gamma+3\right)\left(\left(2\gamma_1\right)-4\gamma+2\right)+4\gamma^2-1\right)}{\left(2\gamma-1\right)\left(2\gamma+1\right)\left(2\gamma+3\right)}\left(\frac{4\times10^{4}}{\zeta}\right)^{3-\gamma}
    \end{align}
\end{appendix}

\end{document}